\newcommand{\be}{\begin{equation}}
\newcommand{\ee}{\end{equation}}

\documentclass[journal,10pt]{IEEEtran}

\usepackage{booktabs}
\usepackage{psfrag}
\ifCLASSINFOpdf
   \usepackage[pdftex]{graphicx}

\else
  
   \usepackage[dvips]{graphicx}

\fi
\usepackage[cmex10]{amsmath}
\usepackage{amsthm,epstopdf,nicefrac,pgfplots}

  \pgfplotsset{compat=newest} 
  \pgfplotsset{plot coordinates/math parser=false}

\usepackage{mathtools}

\usepackage[caption=false,font=footnotesize]{subfig}

\renewcommand\appendix{\par
  \setcounter{section}{0}
  \setcounter{subsection}{0}
  \setcounter{figure}{0}
  \setcounter{table}{0}
  \renewcommand\thesection{Appendix \Alph{section}}
  \renewcommand\thefigure{\Alph{section}\arabic{figure}}
  \renewcommand\thetable{\Alph{section}\arabic{table}}
}

\usepackage{color}

\newcommand\copyrighttext{%
  \footnotesize \textcopyright 2015 IEEE.  Personal use of this material is permitted. Permission from IEEE must be obtained for all other uses, in any current or future media, including reprinting/republishing this material for advertising or promotional purposes, creating new collective works, for resale or redistribution to servers or lists, or reuse of any copyrighted component of this work in other works. 
 DOI: \href{https://doi.org/10.1109/TIM.2015.2390833}{10.1109/TIM.2015.2390833}
 }
\newcommand\copyrightnotice{%
\begin{tikzpicture}[remember picture,overlay]
\node[anchor=south,yshift=10pt] at (current page.south) {\fbox{\parbox{\dimexpr\textwidth-\fboxsep-\fboxrule\relax}{\copyrighttext}}};
\end{tikzpicture}%
}

\usepackage[bookmarks=false]{hyperref}
\twocolumn

\usepackage{balance}
\begin{document}
%
\title{Accurate Measurements using Quantized Data}
\title{Parametric System Identification\\ Using Quantized Data}
%
%
%

\author{A.~Moschitta~\IEEEmembership{Member,~IEEE}\thanks{A. Moschitta is with the University of Perugia - Engineering Department, via G. Duranti, 93 - 06125 Perugia Italy,}
and~J.~Schoukens,~\IEEEmembership{Fellow Member,~IEEE}\thanks{J. Schoukens is with the Vrije Universiteit Brussel, Department ELEC, Pleinlaan 2, B1050 Brussels, Belgium.}
and P.~Carbone~\IEEEmembership{Senior Member,~IEEE}\thanks{P. Carbone is with the University of Perugia - Engineering Department, via G. Duranti, 93 - 06125 Perugia Italy.}}

\newtheorem{theorem}{Theorem}[section]
\newtheorem{lemma}[theorem]{Lemma}

\maketitle
\copyrightnotice
\begin{abstract}
\boldmath
The estimation of signal parameters using quantized data is a recurrent problem in electrical engineering.
As an example, this includes the estimation of a noisy constant value, and of the parameters of a sinewave 
that is its amplitude, initial record phase and  offset. Conventional algorithms, 
such as the arithmetic mean, in the case of the estimation of a constant, are known not to be optimal 
in the presence of quantization errors. They provide biased estimates if particular conditions regarding the quantization 
process are not met, as it usually happens in practice. In this paper a quantile--based estimator is presented that is based on the
Gauss--Markov theorem. The general theory is first described and the estimator is then applied to both DC and AC input 
signals with unknown characteristics. By using simulations and experimental results it is shown that the new estimator
outperforms conventional estimators in both problems, by removing the estimation bias. 

\end{abstract}



\newcommand{\fg}[1]{{\frac{1}{\sqrt{2\pi}\sigma} e^{-\frac{{#1}^2}{2\sigma^2}}}} 

%
\IEEEpeerreviewmaketitle


\section{Introduction}
The estimation of signal parameters based on quantized data is a problem of general interest 
in the area of instrumentation and measurement. 
Frequently the only available information about a physical phenomenon lies in the sequence
of quantized data obtained through an Analog--to--Digital Converter (ADC) and on information
partially available about the input sequence.
As an example, the estimation of a {\color{black}Direct Current} (DC) value, or of the amplitude and the
initial record phase of an {\color{black} Alternate Current} (AC) sequence, fall among such problems: 
samples of the input sequence,
possibly noisy, are converted into digital format for further processing, to identify the needed parameters.  
As shown in \cite{Carbone}\nocite{CarboneSchoukens,Kollarbias, Schuchman,GrayStockham}--\cite{GrayNeuhoff}, unless particular conditions apply, 
the application of conventional algorithms such as the arithmetic mean or the Least Square Estimator (LSE)
result in biased estimates.
Typically, the estimation bias depends on the type of identification problem, e.g. DC or AC type of problem, 
on the noise Probability Density Function (PDF) and on the ADC characteristics.
If the ADC is perfectly uniform, theoretical results can be applied to remove the bias in both the DC and the AC cases, as the quantizer
can be linearized on the average. In practice, however, ADCs are
not uniform. They rather exhibit Integral  (INL) and Differential Nonlinearities (DNL), largely invalidating the hypotheses required for the application of the theoretical results allowing simplified signal processing of quantized samples.  

The problem of identifying signal parameters, after noise is added and quantization is performed, 
can be seen in the larger context of the reconstruction of an input signal PDF: 
in the DC case the PDF is constant over time, in the AC case it becomes time--dependent.
When estimating the PDF at the input of a quantizer through the quantized samples, 
it was proven in \cite{KollarBook} that {\color{black} the following} few cases may occur:
\begin{itemize}
\item if the hypotheses of the {\em quantization theorem} hold true, then the input PDF can be reconstructed with no errors;
\item if the hypotheses of the quantization theorem do not hold true but the noise is Gaussian with a variance comparable to {\color{black} the nominal quantization step} $\Delta$, then the input PDF can be reconstructed by using the results of this theorem with some (negligible) errors;
\item in all other cases, the quantization theorem can not be applied and some other techniques must be used. This occurs{\color{black},} for instance{\color{black},} when the characteristic function of the input PDF is not band--limited, when the input variance is small compared to $\Delta$, or when the quantizer is not uniform, as it happens frequently in practical cases.
\end{itemize} 
When the third case applies, {\em non--subtractive dithering} may {\color{black}partly} relieve from the bias problem, as {\color{black}it} smooths the mean
value of the input--output ADC characteristic 
\cite{GrayStockham}, at the cost of an increased variance.
Alternatively, in solving parametric estimation problems related to the input PDF, 
such as the estimation of the mean of Gaussian noise, a Maximum--Likelihood Estimator (MLE) can be adopted as in \cite{Giaquinto}\nocite{Gendai,CarboneMoschittaSchoukens,Kollar1,Kollar2}--\cite{Gustafsson1}.
However, because of the nonlinearity of the quantizer input--output characteristics, this results in expressions that can be treated only by numerical processing with all practical implications: potential convergence problems, numerical tuning of algorithmic parameters, initialization of the algorithm and local-- instead of global--maxima.

It is shown in this paper that by using  some additional information, e.g. when the noise PDF is known up to a limited set of parameters, ADC data can be used to obtain accurate estimates of unknown parameters, by using linear identification models applied to suitably pre--distorted output samples.
A general approach following this strategy is presented in \cite{WangYinZhangZhao}, where the fundamental underlying theory of system identification based on quantized samples, is presented. A similar approach is followed in this paper where it is shown, both by simulations and experimental results, how to obtain unbiased parametric estimation of the ADC input PDF. 
With respect to \cite{WangYinZhangZhao} the estimator presented in this paper is based directly on the Gauss--Markov theorem and does not require the introduction of a specific procedure leading to the quasi--convex combination estimator. 
Moreover, it estimates the error covariance matrix by avoiding the 
iterated calculations associated {\color{black} with} the recursive approach taken in \cite{104}.
Experimental results obtained by using a commercial Data Acquisition System (DAS) prove the validity of the adopted hypotheses on the simplifying assumptions taken in this paper. 
Its validity is also proven here when the noise standard deviation is small compared to $\Delta$, a typical situation in many cases of practical interest when using ADCs.
The approach is general enough to accommodate for generic noise PDFs and when {\em nuisance} parameters must be estimated along with the input signal parameter values.

Two typical estimation problems are considered in this paper: 
{\color{black}
\begin{itemize}
\item DC case: the estimation of a DC value; 
\item AC case: the estimation of the amplitude, initial record phase and offset of a sinewave, 
\end{itemize}
}
\noindent
when signals are affected by zero--mean Gaussian noise and quantized by a possibly non--uniform ADC.
It is shown that the presented estimator outperforms the arithmetic mean estimator in the DC case and the LSE in the AC case, by largely removing the estimation bias. 

{\color{black}
This paper is organized as follows.
In Section II we introduce the considered system, the associated signals and the adopted
symbol conventions. 
Section III contains the mathematical analysis supporting the
performance of the quantile--based estimator.  
It is organized in subsections to
address both DC and AC estimation problems with the same modeling approach.
Montecarlo--based simulation results are presented in Section IV to validate the derived theory.
Section V contains the description of the experiments done to further prove the estimator properties under the various assumed constraints, while Section VI 
includes comments on the results shown and limits of the proposed estimation procedure. 
}


\section{Signals and systems}

\begin{figure}[b]
\begin{center}
\includegraphics[scale=0.65]{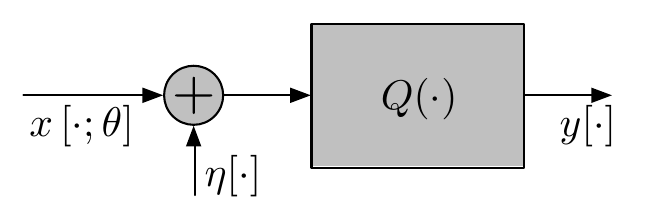}
\caption{The signal chain considered in this paper.\label{signals}}
\end{center}
\end{figure}  
{\color{black} 
In this Section we illustrate 
the properties of the 
quantizing system considered in this paper 
and the type of input signals applied at its input.
The assumed}  signal chain {\color{black} is} 
depicted  in Fig.~\ref{signals}. In this figure, 
$x[\cdot;\theta]$ 
represents a discrete time deterministic sequence known up to a vector parameter $\theta$
and $\eta$ a zero--mean noise sequence with a given PDF {\color{black} and independent outcomes}, whose variance might be unknown. 
The quantizer $Q(\cdot)$  in Fig.~\ref{signals} models the effect of the ADC on the signal. {\color{black} It} might 
be non--uniform, but with $L$ known transition levels,

{\color{black} By assuming $L$ as an even integer,
the quantizer output becomes equal to}
\begin{equation}
{\color{black}
	 y_k  \coloneqq -\left(\frac{L}{2}-1\right) \Delta+ k\Delta, \qquad k=0, \ldots, L-1 
}
\end{equation}
when the input takes values in the interval $[T_{k}, T_{k+1})$, where
$T_k$ is the $k$--th quantizer transition level. Accordingly,
$k=0$ and $k=L-1$ correspond to the quantizer output being equal to $-(\nicefrac{L}{2}-1) \Delta$ and $\nicefrac{L}{2}\Delta$, respectively. 
If the transition levels are unknown, they can be estimated during an initial system calibration phase. {\color{black} It will be shown in Section~\ref{expres} that the estimator
is enough robust to account for uncertainties in the estimated values of the transition levels.} 

{\color{black} It is further assumed that $N$ samples of the ADC output sequence $y[\cdot]$ are collected and processed.}
{\color{black} Consequently,} each ADC output sample {\color{black}can be modeled as} a random variable taking values in $L$ possible categories with probability determined by 
the {\color{black}deterministic} input sequence, the noise PDF and the ADC transition levels.

The additional assumption is made that the quantizer is never overloaded, that is the input signal varies in the input range
that guarantees the quantization error to be granular.
Before describing the proposed estimator{\color{black},} a motivating example is given in the next Section.

\subsection{A motivating example}
\begin{figure}[h]
\begin{center}
\includegraphics[scale=0.45]{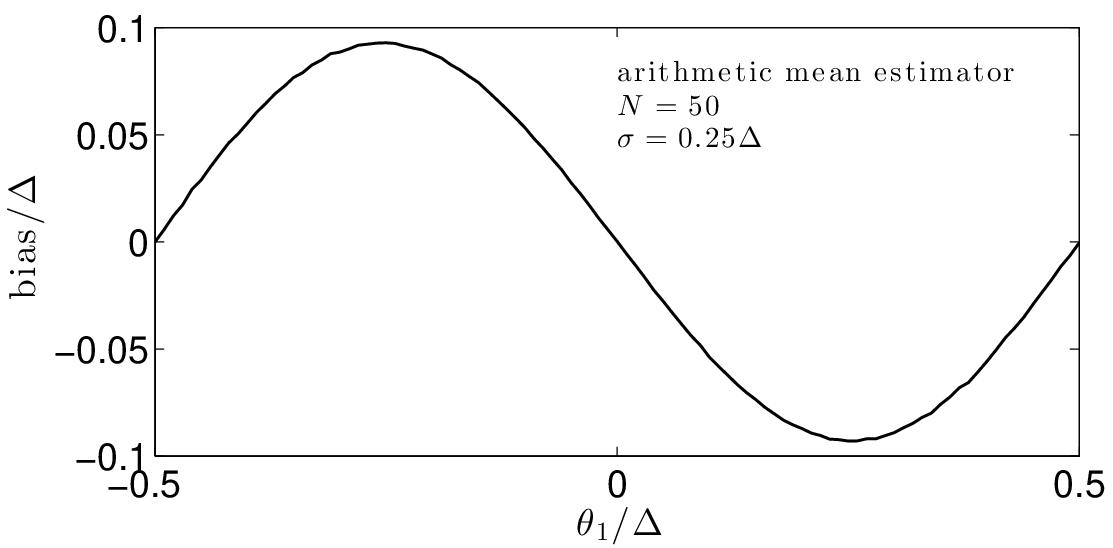}
\caption{Arithmetic mean estimator. Bias in the estimation of a DC value in Gaussian noise with $\sigma=0.25 \Delta$. \label{figbias}}
\end{center}
\end{figure}

Assume that $x[n;\theta]=\theta_1$, that is an unknown constant value.
The natural and most widely used estimator of $\theta_1$ is the arithmetic mean estimator 
\begin{equation}
	\hat{\theta}_1 \coloneqq \frac{1}{N} \sum_{i=0}^{N-1}y[i].
	\label{arithmean}
\end{equation}
If the ADC is uniform and the noise PDF has suitable properties, e.g. the noise characteristic function is band--limited or
appropriate {\em dithering} noise is used, 
$\hat{\theta}_1$ is unbiased. In all other situations that most frequently apply in practice, e.g. when the quantizer inside the ADC is non--uniform or when the noise PDF does not satisfy particular conditions, $\hat{\theta}_1$ is biased.
As an example, assume a $12$-bit ADC uniform in the range $[-1,1)$ and the noise PDF as a zero--mean Gaussian random sequence with $\sigma=0.25\Delta$. 
Under these conditions the bias of (\ref{arithmean}) is shown in Fig.~\ref{figbias}, when $N=50$. 
Since the {\color{black}expected value}  of (\ref{arithmean}) does not actually depend on $N$, even by increasing the number of averaged samples, 
the bias does not vanish. In the following, it will be shown how to remove this bias by using an estimator based on a linear model between data and unknown parameters. 
 

\section{Quantile--based Estimation}
{\color{black} In this Section,} at first the main idea behind the proposed estimator is illustrated using a simple example. 
Then, the {\color{black} parametric} signal models are defined and the full estimator is described in the general case.
\subsection{The estimation of a quantized noisy constant} \label{ss1}
\begin{figure}[t!]
\begin{center}
\includegraphics[scale=0.5]{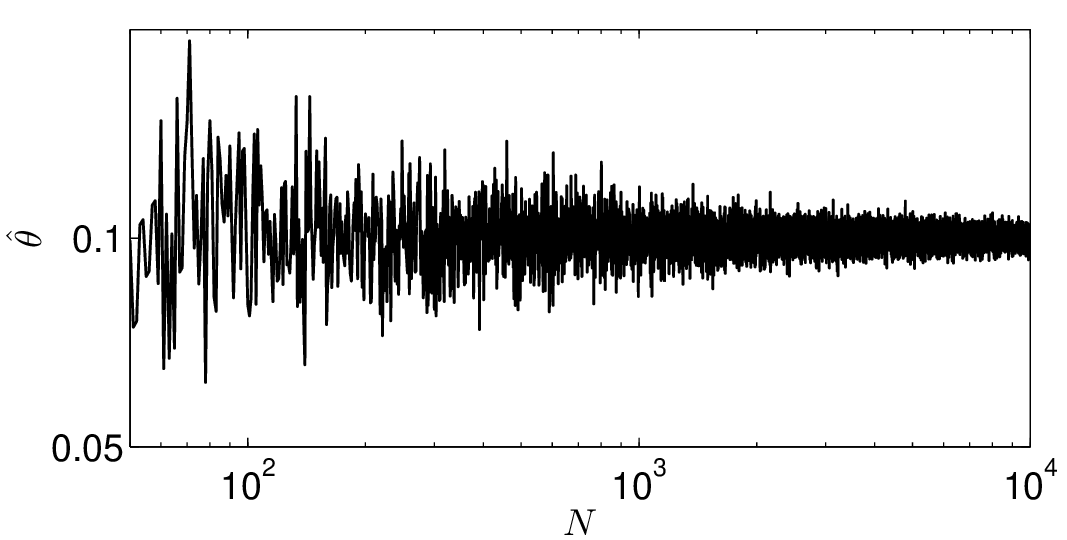}
\caption{Estimate of a constant in noise using {\color{black} (\ref{initest}). Estimator mean value} as a function of the number of samples in $10^1, \ldots, 10^4$: the constant to be estimated is 
$\theta=0.1$, the noise is zero--mean Gaussian with $\sigma=0.1$.}  
\label{figone}
\end{center}
\end{figure}

To show the approach taken in this paper to estimate parametric signals using quantized data{\color{black},} consider the simple 
problem of estimating a constant in noise using a single--bit quantizer, that is a comparator.
Thus,  assume
\begin{equation}
	x[n] = \theta+\eta[n], \quad y[n] = \left\{
		\begin{array}{ll}
			1 & x[n]\ge 0 \\
			0 & x[n]<0
		\end{array}
		\right.
		\quad n=0, \ldots, N-1
\end{equation}
 where $\theta$ is the constant to be identified, $\eta[n]$ is zero--mean Gaussian noise, having a known variance $\sigma^2$ and $y[n]$ is the sequence
 of quantized data. Then simple processing shows that the probability of $y[n]$ being positive is:
 \begin{equation}
 	p_1 \coloneqq P(y[n]=1)=P(x[n] \ge 0) =1-\Phi\left( \frac{-\theta}{\sigma}\right) 
	\label{prob1}
 \end{equation}
Two aspects can be highlighted:
\begin{itemize}
\item by collecting data from the comparator, the probability (\ref{prob1}) can be estimated by elementary processing;
\item when the probability $p_1$ is known, or approximately known, 
(\ref{prob1}) can be inverted to find a value for $\theta$:
\begin{equation}
	\hat \theta =-\sigma \Phi^{-1}\left( 1-\hat{p}_1\right)
	\label{initest}
\end{equation}
\end{itemize}
where $\hat{\theta}$ is an estimator of $\theta$ and $\hat{p}_1$ an estimator of $p_1$.
As an example consider the case $\theta=0.1$, $N=10^1, \ldots, 10^4$. By simple counting the number of times that the comparator outputs {\color{black}equals} $1$ an estimate of $p_1$
is obtained. Then $\theta$ is estimated as shown in Fig.~\ref{figone} as a function of $N$.
This approach can be extended to the case of a multi bit quantizer and to different signal models, as done in the next Sections.

{\color{black} 
\subsection{Extension of the proposed approach to a multi bit quantizer}
 It will be shown that all the 
 estimation problems considered in this paper 
 can be solved by the application of the Gauss--Markov theorem. Accordingly, assume that the sequence of observations $X$
can be  linearly related to the unknown parameters $\theta$ as in the following
\begin{equation}
 X=H\theta + W,
 \label{theform}
\end{equation}
where $X = [x_1\; x_2\; \ldots \; x_N]^T$ represents a column vector containing
outcomes of the observable variable, $\theta=[\theta_1 \; \theta_2\; \ldots \; \theta_M]^T$, the column vector with the unknown parameters to be estimated,
$H$ is a $N \times M$ matrix with known entries and $W=[w_1\;w_2\; \ldots\; w_N]^T$, is a column vector containing outcomes of the noise affecting the observable variables.
Then, if the noise vector is zero--mean, it can be shown that the Best Linear Unbiased Estimator (BLUE) of $\theta$ is \cite{Kay}:
\begin{equation}
	\hat\theta \coloneqq (H^T\Sigma_{X}^{-1}H)^{-1}H^T\Sigma_{X}^{-1}X,
	\label{BLUEdef}
\end{equation} 
where $\Sigma_X$ represents the covariance matrix of  the noise vector $W$.
In the next subsections it is shown how to cast several parametric identification models in the form (\ref{theform}), so to apply (\ref{BLUEdef}) for their solution. The added complexity here refers to the effect of quantization that correlates outcomes and distorts input data because of its nonlinear input--output characteristic.   
Thus, each problem considered in the following will be addressed by:
\begin{itemize}
\item considering the effect of quantization and showing how to linearize the relationship between observable data (the quantized output sequence) and unknown parameters, as
in (\ref{theform});
\item  showing that the quantizer output sequence provides useful information for the unbiased estimation of the input noise quantiles; 
\item illustrating how to estimate the noise covariance matrix required by (\ref{BLUEdef}), by using available information provided by the quantizer;
\item using (\ref{BLUEdef}) to provide an expression for the estimator in the considered cases. 
\end{itemize}
{\color{black} In subsections~C--H the idea presented in subsection~\ref{ss1} is extended to comprehend the case of several quantization levels. At first, the considered 
DC and AC models are presented in subsection~C and then the 
estimator forms are expressed, by additionally assuming a Gaussian noise PDF.}
}

{\color{black}
\subsection{The DC and AC parametric signal models}}
{\color{black} In this subsection, the DC and AC data models considered in the following subsections are presented.}
Assuming $n=0, \ldots, N-1$, {\color{black}the  analyzed DC parametric models  are}:
\begin{align}
\begin{split}
	& \mbox{model \;$1$: \;} y[n] = \theta_1+\sigma\eta[n]+e[n], \\
	& \mbox{model \;$2$: \;} y[n] = \theta_1+\theta_2\eta[n]+e[n],
\end{split}
\end{align}
where $e[\cdot]$ is the quantization error sequence. While in model~1
the noise standard deviation $\sigma$ is assumed to be known, in model~2 this is
taken as a second parameter to be estimated.
The identification of a third AC parametric model, 
\begin{align}
	\mbox{model \;$3$: \;} y[n] = \theta_0+\theta_1x_1[n]+\theta_2x_2[n] +\sigma \eta[n]+e[n],
	\label{dynmodel}
\end{align}
is considered, where $x_1[n]$, $x_2[n]$ are known periodic sequences and $\theta_i$,
$i=0,1,2$, are three parameters to be identified.
The additional assumption for this model is that sampling is {\em coherent}, that is {\em synchronous} and
such that $N$ corresponds to an integer number of periods of both $x_1[\cdot]$ and $x_2[\cdot]$.
For instance, if $x_1[n]=\cos(\omega n)$ and $x_2[n]=\sin(\omega n)$ 
where $\omega$ is a known constant,
(\ref{dynmodel}) represents the well known model of the three--parameter sine fit \cite{IEEE1241}.
In the following, the properties of a quantile--based estimator applied
to all these cases will be illustrated.

\subsection{The ADC as a Source of Ordinal Data}
{\color{black} In this subsection, the statistical properties of the sequence of data
output by an ADC are recalled, so to serve as a basis for the proposal
of a quantile--based estimator.}
As a general remark, the quantizer inside the ADC maps the input values to an ordinal scale 
that admits calculation of the mean value of the measurement results, as in (\ref{arithmean}), 
not without controversy\footnote{As pointed out by \cite{Knapp}, even Stevens makes a practical concession to the usage
of otherwise not admissible statistics of ordered data \cite{Stevens}: {\em In the strictest propriety 
the ordinary statistics involving means and standard deviations ought not to be used with these (ordinal)
scales, for these statistics imply a knowledge of something more than the relative rank--order of data. On the other hand, for this 'illegal' statisticizing there can be invoked a kind of pragmatic sanction: In numerous instances it leads to fruitful results.}}.
Conversely, generally applicable statistics include the estimation of quantiles that will be applied in the following to remove 
the potential incongruences associated {\color{black}with} the usage of (\ref{arithmean}).
In this subsection an analysis is made on the properties of the information available when
solving model~1 through model~3 problem types. 

For a given value of $\theta_1$ in model~1 and model~2  define,
$
	\Pi =\left[ p_{0} \ldots p_{L-1}  \right]^T
$
where each $p_k$ represents the probability of $y[\cdot]$ taking the value $y_k$, 
and
$
	C =\left[ c_{0} \ldots c_{L-1}  \right]^T
$
where $c_k \coloneqq Np_k$, is the average number of occurrences in code bin $k$ when $N$ samples are collected.
Also define, 
$
	C\Pi = [cp_0 \ldots cp_{L-1}]^T 
$
where $cp_k = \sum_{n=0}^{k}p_n$.
Moreover define 
$
	\hat{C} =\left[ \hat{c}_{0}\; \ldots \; \hat{c}_{L-1}  \right]^T
$
as the random vector containing the {\em experimental} number of occurrences in each code bin 
and 
$
	\widehat{C\Pi}  = [\widehat{cp}_0 \ldots \widehat{cp}_{L-1}]^T
$
where $\widehat{cp}_k \coloneqq \sum_{n=0}^{k}\hat{c}_n$.

Then $\hat{C}$ is a random variable having {\color{black} a}  multinomial distribution with parameters $N$ and $\Pi$ for which \cite{multinomial}:
\begin{equation}
	E(\hat{C}) = 
	\left[
	\begin{array}{c}
	Np_{0} \\
	\vdots \\
	Np_{L-1} \\
	\end{array}
	\right]
\end{equation}
and whose covariance matrix is {\color{black} \cite{multinomial}}:
\begin{equation}
	\Sigma_{\hat{C}} =
	\left[
		\begin{array}{cccc}
		Np_0(1-p_0) & -Np_0p_1 & \ldots &   -Np_0p_{L-1} \\
		-Np_0p_1  & Np_1(1-p_1) & \ldots &   -Np_1p_{L-1} \\
		\vdots & \vdots & \vdots & \vdots \\
		-Np_0p_{L-1}  & -Np_1p_{L-1}   & \ldots &   Np_{L-1}p_{L-1} \\
		\end{array}
	\right]
	\label{mc}
\end{equation}
Then the maximum likelihood estimator of $p_k$ is, 
\begin{equation}
	\hat{p}_k = \frac{\hat{c}_k}{N} \qquad k =0, \ldots, L-1 
	\label{phat}
\end{equation}
and $\hat \Pi =[\hat{p}_0\; \ldots\; \hat{p}_{L-1}]^T$ is an unbiased estimator of $\Pi$ with covariance matrix $\Sigma_{\hat{\Pi}} =\frac{1}{N^2} \Sigma_{\hat{C}}$.

{\color{black}Observe that regardless of the {\em true} value of the sequence generating a given quantized output sequence, when the problem is a static one, the only available information at the quantizer output can be modeled as done in this subsection. Thus, the open problem remains that of exploiting efficiently these data 
to extract all possible information about the unknown parameter. Moreover,}
since off--diagonal entries in the covariance matrix $\Sigma_{\hat{C}}$ are not null, it is expected that the simple mean estimator of $\theta_1$ will not yield optimal statistical performance, as it ignores both that different values have different probability of occurrence and the mutual information carried by different values of the quantized sequence. Alternative estimators can be 
employed as shown in the following.
 
\subsection{Model 1}
By assuming that the quantizer is not overloaded we have:
\begin{align}
\begin{split}
	F_y(y_k) \coloneqq & P(y[n] \leq y_k)  = P \left( \theta_1 +\sigma\eta[n] \leq T_k \right) \\ 
	= & F\left(\frac{T_k-\theta_1}{\sigma}\right) \\
	= &\sum_{n=0}^{k-1} p_n 	\coloneqq cp_k,
	\quad k=1, \ldots, L-1 
	\label{thirteen}
\end{split}
\end{align}
with $F(\cdot)$ as the noise cumulative distribution function, from which we derive
\begin{equation}
	T_k = F_y^{-1}(cp_k)  = \theta_1+\sigma F^{-1}(cp_k) \quad k=1, \ldots, L-1.
	\label{quantest}
\end{equation}
Since $cp_k$ can be estimated using
experimental data, (\ref{quantest}) is the key equation for the proposal of a quantile--based estimator \cite{WangYinZhangZhao}.
{\color{black}While a mathematical form of the 
estimator could be derived by resorting to the case of a generic
noise PDF, the following subsection adds the hypothesis of Gaussian noise to reduce the level of abstraction and to increase usability of results. 
The approach taken in the next subsection under the assumption 
of model~1 will be then extended in a similar way to comprehend also model~2 and model~3.}

\subsection{Gaussian case}
In the Gaussian case, we have $F(x)=\Phi\left( x\right)$, where $\Phi(\cdot)$ is the 
{\color{black} cumulative distribution function}
of a standard Gaussian random variable,
so that from (\ref{quantest}) 
\begin{equation}
	{\theta}_1 = T_k-\sigma \Phi^{-1}({cp}_k) \qquad k=1, \ldots, L-1. 
	\label{linearmodel}
\end{equation}
that shows that there is a linear relationship 
between suitably pre--distorted {\color{black} cumulative probabilities defined in (\ref{thirteen}), 
and the constant input $\theta_1$. To derive an expression 
for the estimator of $\theta_1$, 
$cp_k$ will be substituted by $\hat{cp}_k$.} 
{\color{black}Accordingly,} define
 ${\cal L} =\{k_1, \ldots, k_{\Lambda} \}$ as the set of indices in the interval $0, \ldots, L-1$ for which $0 < {\hat{p}}_k< 1$
 and $\Lambda \ge 1$ its cardinality and
\begin{align}
\begin{split}
	& 
	H_1 \coloneqq [  \overbrace{1  \; \cdots \; 1 }^{\Lambda}]^T \\
	& 
	X_1 \coloneqq \left[T_{k_1}-\sigma \Phi^{-1}(\widehat{cp}_{k_1}) \; \cdots \; T_{k_\Lambda}-\sigma \Phi^{-1}
(\widehat{cp}_{k_\Lambda}) \right]^T\\
	&
	W_1 \coloneqq X_1-H_1\theta_1,  	
\end{split}
\end{align}
where $W_1$ is a noise vector having covariance matrix $\Sigma_{X_1}$.
The estimation problem can then be written in matrix form:
\begin{equation}
	X_1=H_1\theta_1 + W_1
	\label{matriceq}
\end{equation}
When $N$ is sufficiently large, the variance of $\widehat{cp}_k$ is sufficiently small to allow usage of a first--order Taylor series expansion of each component in $X_1$
about  $E( \widehat{cp}_k )=cp_k$.
As a consequence, the nonlinear function $\Phi^{-1}(\cdot)$ is linearized so that  $E(W_1) \simeq 0$ results and (\ref{matriceq}) satisfies the hypotheses
of the Gauss--Markov theorem. Accordingly, the {\color{black} BLUE} of $\theta_1$ can be written as \cite{Kay}:
\begin{equation}
	\hat\theta_{GM_1} \coloneqq (H_1^T\Sigma_{X_1}^{-1}H_1)^{-1}H_1^T\Sigma_{X_1}^{-1}X_1
	\label{BLUE}
\end{equation}
with
\be
	\mbox{Var}\left( {\hat{\theta}_{GM_1}} \right)= (H_1^T\Sigma_{X_1}^{-1}H_1)^{-1}
\ee
where $\Sigma_{X_1}$ also represents the covariance matrix of $X_1$. 

An estimate for the covariance matrix of $X_1$ is obtained by observing that:
\begin{equation}
	{\Sigma}_{X_1} = \sigma^2 \Sigma_{Y},
\end{equation}
where $Y = \left[\Phi^{-1}(\widehat{cp}_{k_1})\; \cdots \; \Phi^{-1}(\widehat{cp}_{k_\Lambda}) \right]^T$ and
$\Sigma_Y$ its covariance matrix.
An approximated expression for $\Sigma_Y$ can be obtained by linearizing
the nonlinear function $\Phi^{-1}(\cdot)$ using a Taylor series expansion about
the mean value of each component in $\widehat{C\Pi}$, so that we can write {\color{black}\cite{Kai}}:
\begin{equation}
	\Sigma_Y \simeq J \Sigma_{\widehat{C\Pi}} J^T,
		\label{jacob}
\end{equation}
where $J$ is a diagonal matrix defined as
\begin{equation}
	J = \mbox{diag}
	\left(
	\left. \frac{d \Phi^{-1}(x)}{dx}\right|_{x=\widehat{cp}_{k_1}}, \ldots, 
	\left. \frac{d \Phi^{-1}(dx)}{dx}\right|_{x=\widehat{cp}_{k_\Lambda}} 
	\right).
\end{equation}
By recalling that the derivative of the inverse function can be expressed in terms of the derivative of the 
direct function, we have:
\begin{align}
\begin{split}
	\left.
		\frac{d\Phi^{-1}(x)}{dx}
	\right|_{x=\widehat{cp}_k} 
	= 
	\frac{1}
	{\left. \frac{d\Phi(x)
		}{dx}
		\right|_{x=\Phi^{-1}(\widehat{cp}_k)} 
	}   
	= &
	\sqrt{2\pi}e^{\frac{1}{2}\left( \Phi^{-1}(\widehat{cp}_k)\right)^2} \\
	& k=k_1, \ldots, k_\Lambda
\end{split}
\end{align}
Finally, an estimate of the covariance matrix of $\widehat{C\Pi}$ to be substituted in (\ref{jacob}),
can be obtained by observing that  ${C\Pi}=A\Pi$
where
\begin{equation}
	A=
	\left[
	\begin{array}{cccc}
	1 & 0 & \cdots & 0 \\
	1 & 1 & \cdots & 0 \\
	\vdots & \vdots & \cdots &\vdots \\
	1 & 1 & \cdots & 1
	\end{array} 	
	\right]
\end{equation}
is a lower diagonal matrix. 
Thus the covariance matrix of $\widehat{C\Pi}$ is 
\begin{equation}
	\Sigma_{\widehat{C\Pi}} = A\Sigma_{\hat \Pi}A^T,
\end{equation}
and estimates of $\Sigma_{\hat \Pi}$ and $\Sigma_{\widehat{C\Pi}}$ are obtained by 
replacing each entry in (\ref{mc}) of the type $Np_ip_j$, by the corresponding natural 
estimator based on the product of estimated probabilities $N\hat{p}_i\hat{p}_j$, with
$\hat{p}_i$ defined in (\ref{phat}). Once all unknown quantities are substituted by their estimates in (\ref{BLUE}), a similar 
estimator to that presented in \cite{WangYinZhangZhao} (e.g. eq. 6.19) is obtained.
\subsection{Model 2}
By considering model 2 under the hypothesis of Gaussian noise, the equivalent expression for (\ref{quantest}) is:
\begin{equation}
	T_k = F_y^{-1}(cp_k)  = \theta_1+\theta_2 F^{-1}(cp_k) \quad k=1, \ldots, L-1
	\label{quantestt}
\end{equation}
from which, {\color{black} using the Gaussian hypothesis,} we have:
\begin{equation}
	\frac{T_k}{\theta_2}-\frac{{\theta}_1}{\theta_2} = \Phi^{-1}({cp}_k) \qquad k=1, \ldots, L-1. 
	\label{linearmodell}
\end{equation}
To linearize the relationship between parameters and observed data, 
define a new vector parameter $\gamma \coloneqq [\gamma_1 \; \gamma_2]^T \coloneqq \left[ \frac{1}{\theta_2} \; \frac{\theta_1}{\theta_2} \right]^T$. Observe that if an estimate for $\gamma$ is available, an estimate for $\theta_1$ and $\theta_2$ is easily obtained by inverting the relationship in the definition of $\gamma$. 
Thus (\ref{linearmodell}) can be rewritten as:
\begin{equation}
T_k\gamma_1-\gamma_2 = \Phi^{-1}(cp_k) \qquad k=1, \ldots, L-1.
\end{equation}
Then again define
 ${\cal L} = \{ k_1, \ldots, k_\Lambda\}$ as the set of indices in the interval $0, \ldots, L-1$ for which $0 < {\hat{p}}_k< 1$,
 $\Lambda \ge 2$ its cardinality and
\begin{align}
\begin{split}
	& 
	H_2 \coloneqq 
	\left[ 
		\begin{array}{cc}
			T_{k_1} & -1 \\
			T_{k_2} & -1 \\
			\vdots & \vdots \\
			T_{k_\Lambda} & -1  \\
		\end{array}
	\right]\\
	& 
	X_2 \coloneqq \left[\Phi^{-1}(\widehat{cp}_{k_1}) \; \cdots \;  \Phi^{-1}
(\widehat{cp}_{k_\Lambda}) \right]^T\\
	&
	W_2 \coloneqq X_2-H_2\gamma,  	
\end{split}
\end{align}
where $W_2$ is a noise vector having covariance matrix $\Sigma_{X_2}=\Sigma_Y$.
The estimation problem can then be written in matrix form:
\begin{equation}
	X_2=H_2\gamma + W_2
	\label{matriceqq}
\end{equation}
Again, when $N$ is sufficiently large, each component in $X_2$ can be linearized about the corresponding mean value
as in the case of model~1 and $E(W_2) \simeq 0$ results.
The application of the Gauss--Markov theorem provides:
\begin{equation}
	\hat\theta_{GM_2} \coloneqq (H_2^T\Sigma_{X_2}^{-1}H_2)^{-1}H_2^T\Sigma_{X_2}^{-1}X_2
\end{equation}
with
\be
	\mbox{Var}\left( {\hat{\theta}_{GM_2}} \right)= (H_2^T\Sigma_{X_2}^{-1}H_2)^{-1}
\ee
where $\Sigma_{X_2}$ also 
represents the covariance 
matrix of $X_2$ and is estimated as in the case of model~1.

\subsection{Model 3}
\begin{figure}[b!]
\begin{center}
\includegraphics[scale=0.5]{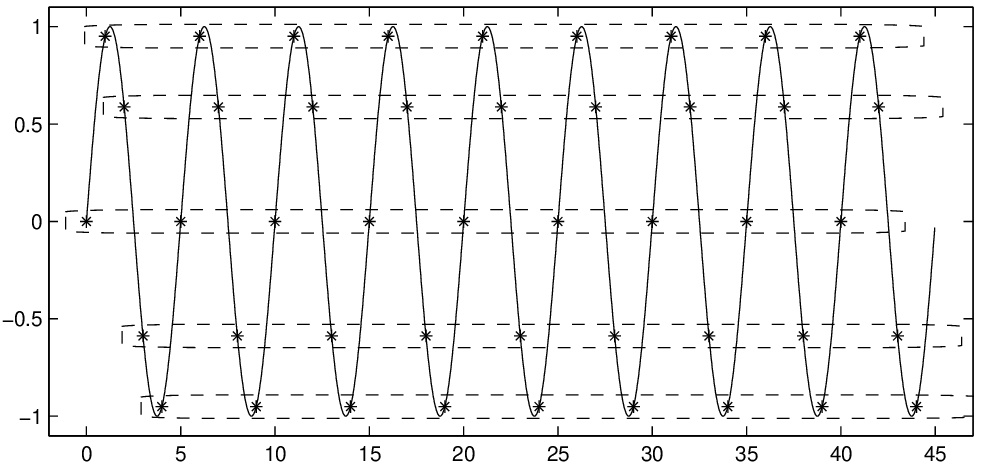}
\caption{Model 3.  Sinewave coherent sampling in absence of additive noise and quantization: 
$N=9$ periods of the sinusoidal signal having each $M=5$ samples per period. 
$N$ independently sampled values are obtained that are periodic with period $M$.
\label{figsinfit}}
\end{center}
\end{figure}

The coherency condition on sampling implies that an integer number $N$ of periods of $x_1[\cdot]$
and $x_2[\cdot]$ are observed. Let us further assume that each period contains $M$ samples of the input signal.
Consequently, the total number of samples is $K=MN$ 
and 
{\color{black} if 
the noise $\eta[\cdot]$ would not be present}, the quantizer
output would be periodic with the same period $P=MT_S$ of the known sequences, with  
$T_S$ as the sampling period.
By recalling that any real number $x$ can be expressed as the sum of its integer part $\lfloor x \rfloor$
and of its fractional part $\langle x \rangle$ we can write
$$
x_i[n] = 
x_i\left[ \left \lfloor \frac{n}{M} \right \rfloor M +\left \langle \frac{n}{M} \right\rangle M\right]
= x_i\left[ \left \langle \frac{n}{M} \right\rangle M\right] \; i=1,2,
$$
where the equality follows by the periodicity assumption.
Thus, there are only $M$ 
different time instants {\color{black} modulo $M$} associated {\color{black} with} $n=0, \ldots, M-1$, 
each one recorded $N$ times, that is the number of periods. 
This is made clear in Fig.~\ref{figsinfit} where a sinusoidal sequence $x_1[n]$ is assumed with $M=5$ and $N=9$.
Therefore a total of $K=45$ samples is collected, of which only $5$ refer to independent time instants.
Dashed rectangles in this figure show the samples that provide the same signal amplitude, when
the effects of noise and quantization are neglected. 
Now consider a given value of $n \in \{0, \ldots, M-1\}$ and
$m \in I_n \eqqcolon\{ n, n+M, n+2M, \ldots, n+(N-1)M\}$. Since, for some integer $i$,
$$
\left \langle \frac{m}{M} \right \rangle M = \left \langle \frac{n+iM}{M} \right \rangle M =
\left \langle \frac{n}{M} \right  \rangle M = \frac{n}{M}M = n   
$$
for $m \in I_n$ we can write
\begin{equation}
	y[m] = \theta_0+\theta_1x_1[n]+\theta_2x_2[n] +\sigma \eta[m]+e[m],
	\label{sinseq}
\end{equation}
that provides $N$ realizations of the quantizer output for any single value of the constant input 
$\theta_0+\theta_1x_1[n]+\theta_2x_2[n]$. 
Thus, $I_0, \ldots, I_{M-1}$ provide a partition of the $K$ samples.
Moreover,  for a given value $n$,  (\ref{sinseq})
provides $N$ values that can be used to build a histogram of the quantized output, 
\begin{equation}
	\hat{C}[n] =\left[ \hat{c}_{0}[n]\; \ldots \; \hat{c}_{L-1}[n]  \right]^T
\end{equation}
as the random vector containing the {\em experimental} number of occurrences in each code bin 
when the deterministic input is $s[n] \coloneqq \theta_0+\theta_1x_1[n]+\theta_2x_2[n]$ and,
$
	\widehat{C\Pi}[n]  = [\widehat{cp}_0[n] \ldots \widehat{cp}_{L-1}[n]]^T
$
where $\widehat{cp}_k[n] \coloneqq \sum_{n=0}^{k}\hat{c}_n[n]$.

By considering model 3 under the hypothesis of Gaussian noise, the equivalent expression for (\ref{quantest}) is:
\begin{align}
\begin{split}
	F_y(y_k) \coloneqq & P(y[n] \leq y_k)  \\
	= & P \left( \theta_0+\theta_1x_1[n]+\theta_2x_2[n] +\sigma\eta[n] \leq T_k \right) \\ 
	= & \Phi\left(\frac{T_k-\theta_0-\theta_1x_1[n]-\theta_2x_2[n] }{\sigma}\right) \\
	= &\sum_{h=0}^{k-1} p_h[n] 	\coloneqq cp_k[n],
	\quad k=1, \ldots, L-1
	\label{mod3} 
\end{split}
\end{align}
where $p_h[n]$ represents the probability that $y[m]$ takes the value $h$ when the input signal
is $s[n]$. 
Thus, from (\ref{mod3}) we have:
\begin{align}
\begin{split}
\theta_0+  \theta_1 x_1[n]+ \theta_2 x_2[n] & =  T_k-  \sigma \Phi^{-1}\left( cp_k[n] \right)  \\
 & k=1,\ldots, L-1 \quad
n=0, \ldots, M-1
\end{split}
\end{align}
For each $n$, define a corresponding set ${\cal L}[n]$ as the set of indices in the interval $0, \ldots, L-1$
for which $0 < \hat{p}_k[n] < 1$ and $\Lambda[n]$ its cardinality.
Then define
\begin{align}
\begin{split}
	& 
	\theta_3\coloneqq [\theta_0 \; \theta_1\; \theta_2]^{T} \\
	&
	H_3[n] \coloneqq 
	\left[ 
		\begin{array}{ccc}
			1 & x_1[n] & x_2[n]  \\
			1 & x_1[n] & x_2[n]\\
			\vdots & \vdots & \vdots \\
			1 & x_1[n]  & x_2[n]\\
		\end{array}
	\right]\\
	& 
	X_3[n] 
	\coloneqq 
	\left[ 
		\begin{array}{c}
			T_1 -\sigma \Phi^{-1}(\widehat{cp}_1[n])  \\
			T_2 -\sigma \Phi^{-1}(\widehat{cp}_2[n])\\
			\vdots \\
			T_{\Lambda[n]}-\sigma\Phi^{-1}(\widehat{cp}_{\Lambda}[n])  \\
		\end{array}
	\right]\\
	&
	W_3[n] \coloneqq X_3[n]-H_3[n]\theta_3. 	
\end{split}
\label{h3n}
\end{align}
Then, an estimate of the covariance matrix of $W_3[n]$ is 
$\Sigma_{W_3}[n] 
\coloneqq \sigma^2\Sigma_Y[n]$, where $\Sigma_Y[n]$
is based on the definition in (\ref{jacob}), in which each occurrence of $\widehat{cp}_k$ is substituted by $\widehat{cp}_k[n]$.
Finally, the matrices for the entire set of $M$ time--dependent input values {\color{black} are} constructed by defining:
\begin{align}
\begin{split}
	& 
	H_3 \coloneqq \left[ H_3[0]^T  \cdots H_3[M-1]^T \right]^T \\
	&
	X_3 \coloneqq \left[ X_3[0]^T\; X_3[1]^T\; \cdots \; X_3[M-1]^T \right]^{T} \\
	& 
	\Sigma_{X_3}  \coloneqq \mbox{diag}\left( {\Sigma_{W_3}[0], \ldots, \Sigma_{W_3}[M-1]} \right)^{T} 
\end{split}
\label{h3}
\end{align}
Then the application of the Gauss--Markov theorem provides:
\begin{equation}
	\hat\theta_{GM_3} \coloneqq (H_3^T\Sigma_{X_3}^{-1}H_3)^{-1}H_3^T\Sigma_{X_3}^{-1}X_3
\end{equation}
with
\be
	\mbox{Var}\left( {\hat{\theta}_{GM_3}} \right)= (H_3^T\Sigma_{X_3}^{-1}H_3)^{-1}
\ee
Observe that:
\begin{itemize}
\item if for some $n$, 
a single quantization bin is excited by the corresponding 
quantizer input sample, then $H_3[n]$ in (\ref{h3n}) vanishes, it must be discarded and not included in the dataset used to form $H_3$ in (\ref{h3});
\item enough information is needed to estimate the $3$ scalar parameters in $\theta_3$, 
that is the number of rows of $H_3$ in (\ref{h3}) must be not less than $3$.
\end{itemize}
{\color{black} In the two following Sections both simulation and experimental results are presented.}
\begin{figure}[t!]
\begin{center}
\includegraphics[scale=0.35]{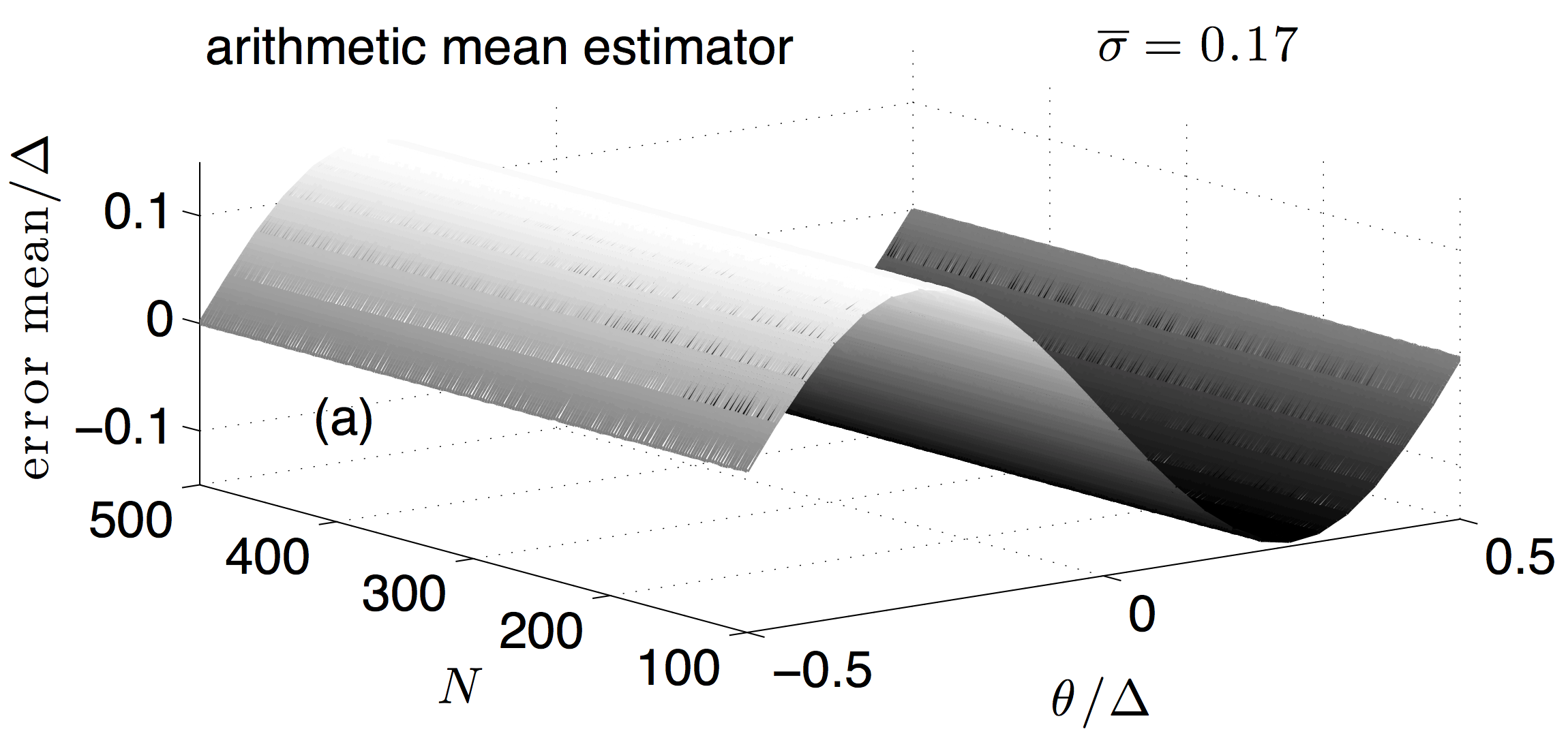}  
\includegraphics[scale=0.35]{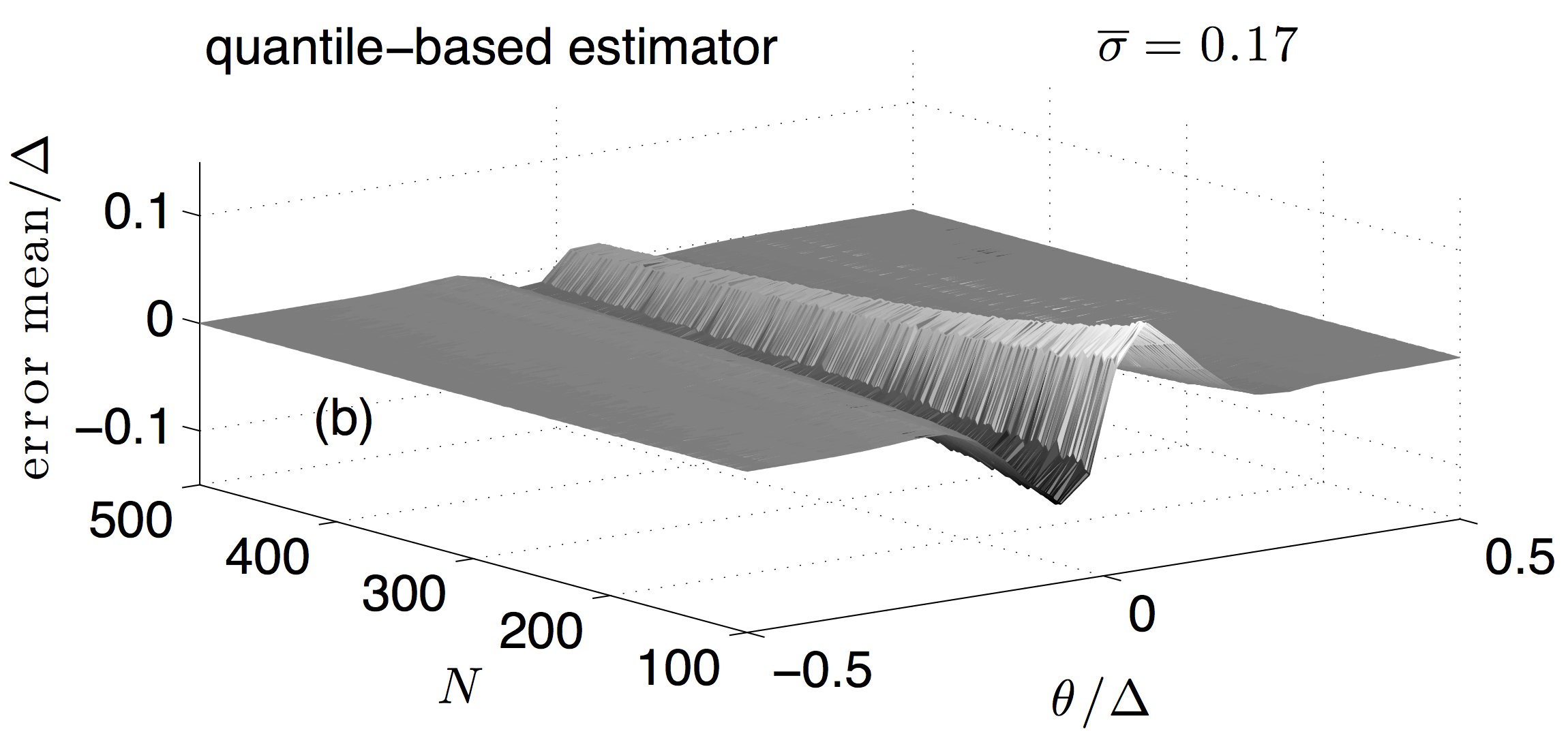}
\includegraphics[scale=0.35]{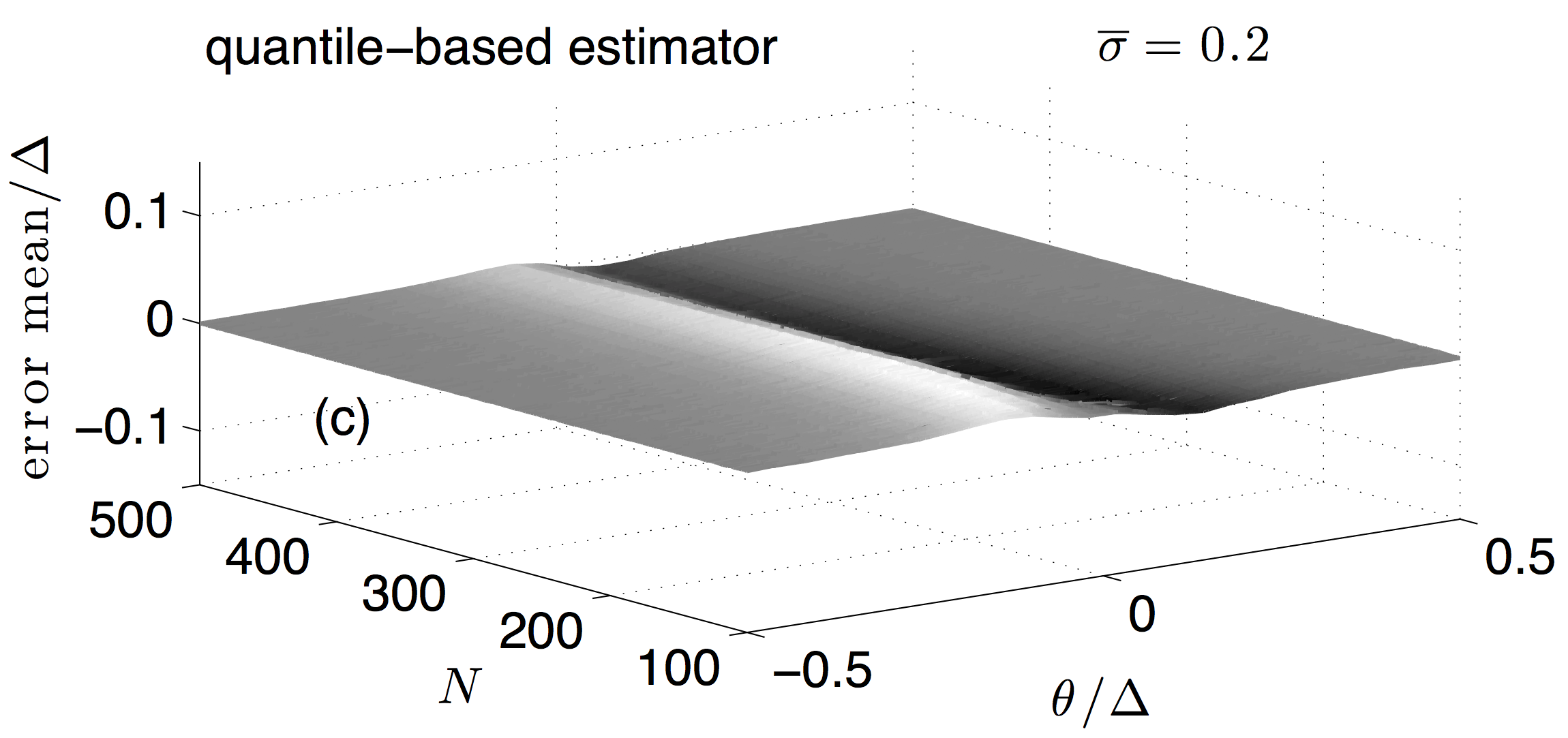}
\caption{Model 1. Mean value of the estimation error of a quantized constant in noise normalized to $\Delta$. Montecarlo results based on $5000$ records as a function of $\nicefrac{\theta}{\Delta}$ and $N$: arithmetic mean estimator (a), quantile--based estimator 
when $\overline{\sigma}=0.17$ (b) and $\overline{\sigma}=0.2$ (c).  
\label{errmone}}
\end{center}
\end{figure}

\section{Simulation results}
\begin{figure}[t!]
\begin{center}
\includegraphics[scale=0.35]{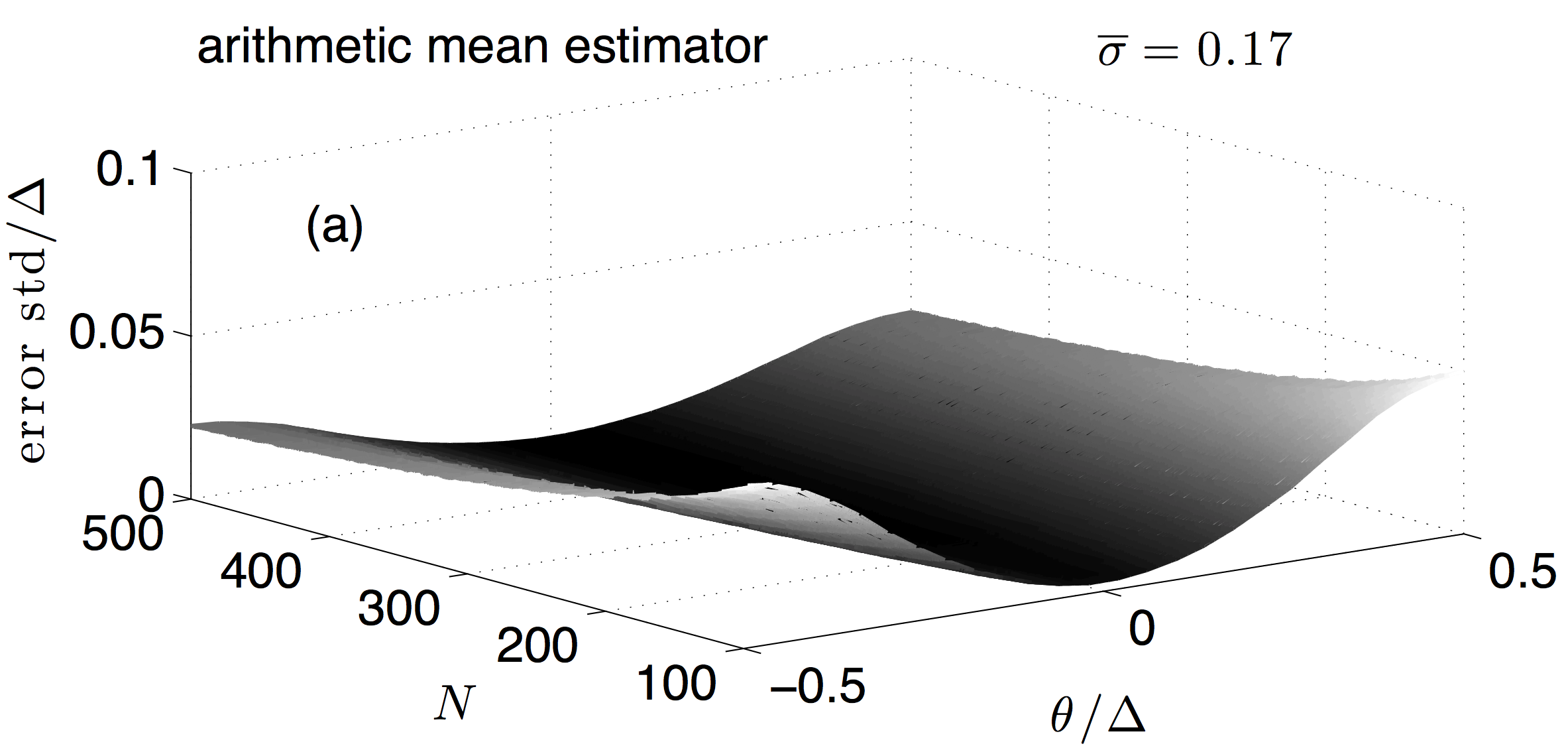}  
\includegraphics[scale=0.35]{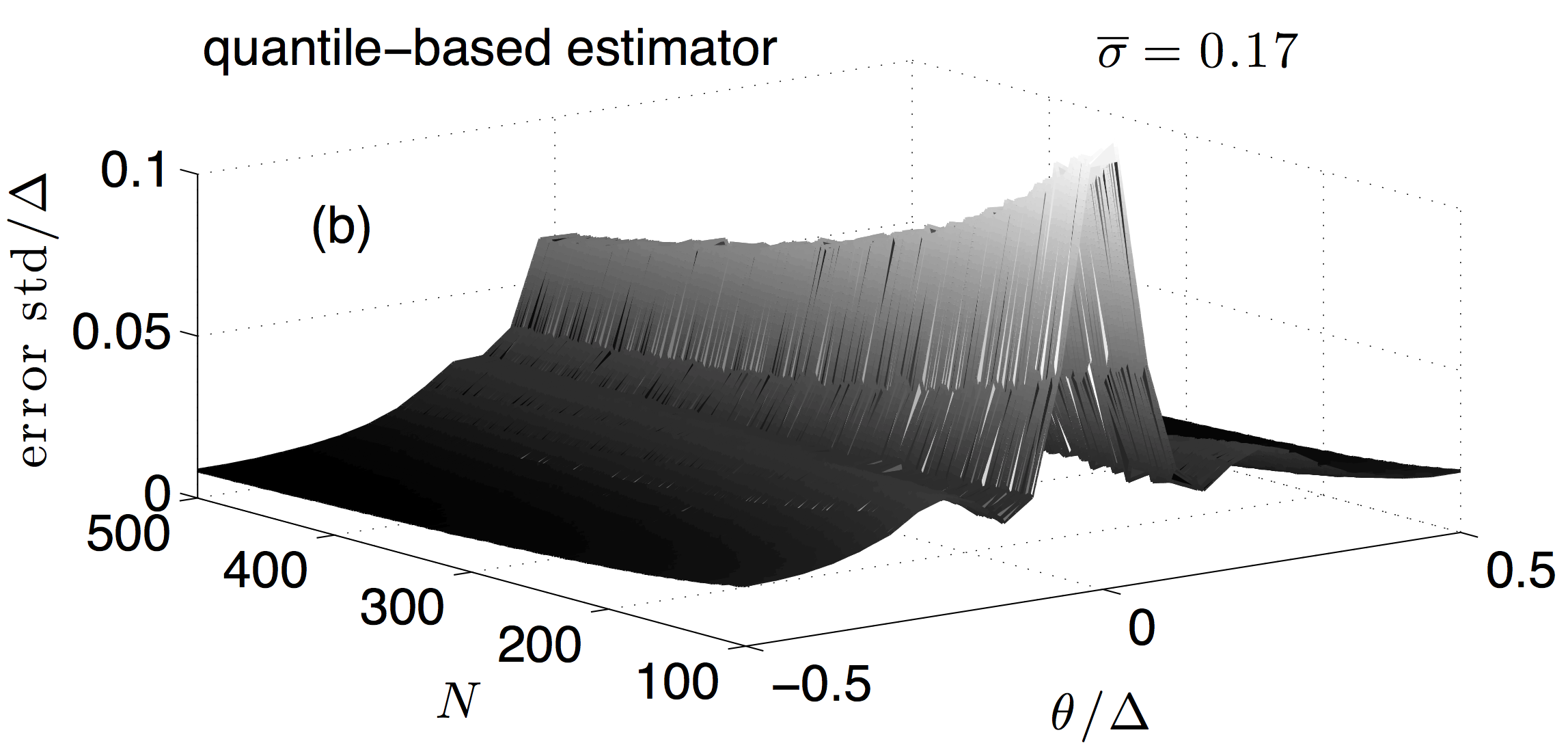}
\includegraphics[scale=0.35]{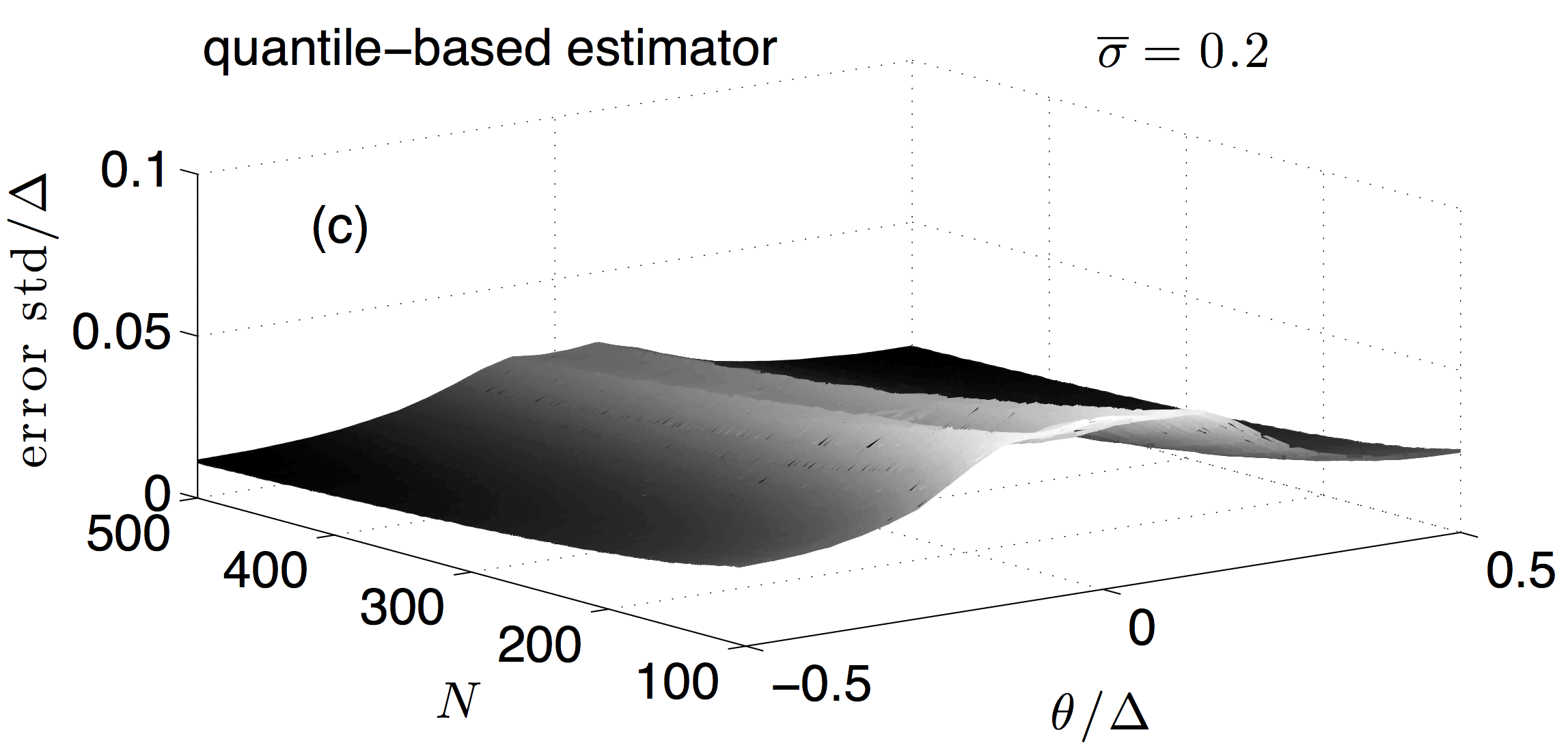}
\caption{Model 1. Standard deviation of the estimation error of a quantized constant in noise normalized to $\Delta$. Montecarlo results based on $5000$ records as a function of $\nicefrac{\theta}{\Delta}$ and $N$: arithmetic mean estimator (a), quantile--based estimator
when $\overline{\sigma}=0.17$ (b) and $\overline{\sigma}=0.2$ (c).  
\label{errsone}}
\end{center}
\end{figure}

At first model~1 was considered.
Mean values and standard deviations of the estimation errors obtained by the Montecarlo approach based on $R=5000$ records are presented in Fig.~\ref{errmone} and in Fig.~\ref{errsone}, respectively,
as a function of both $\nicefrac{\theta}{\Delta}$ and $N=100, \ldots, 500$.
{\color{black}
All results were obtained assuming $\Delta=2/2^b$, where
$b=10$ is the number of bits, and were normalized with respect to $\Delta$.}
Given the periodic behavior of the quantization error input--output characteristic, curves in Fig.~\ref{errmone}
and Fig.~\ref{errsone} are periodic with $\Delta$ and their behavior is shown here, assuming the single
period $-\nicefrac{\Delta}{2} < \theta < \nicefrac{\Delta}{2}$. 
The estimation mean error is plotted in Fig.~\ref{errmone} in the case of the simple arithmetic mean (a) and
in the case of the quantile--based estimator (b-c). While the simple arithmetic mean estimation error does not depend on 
$N$,
the performance of the newly proposed estimator also depends on the number 
of averaged samples. This is a common behavior in the case of bias--removing procedures \cite{CarboneVandersteen}.
In both cases, the mean square error decreases when increasing the number of processed samples.
 
\begin{figure}[t!]
\begin{center}
\includegraphics[scale=0.4]{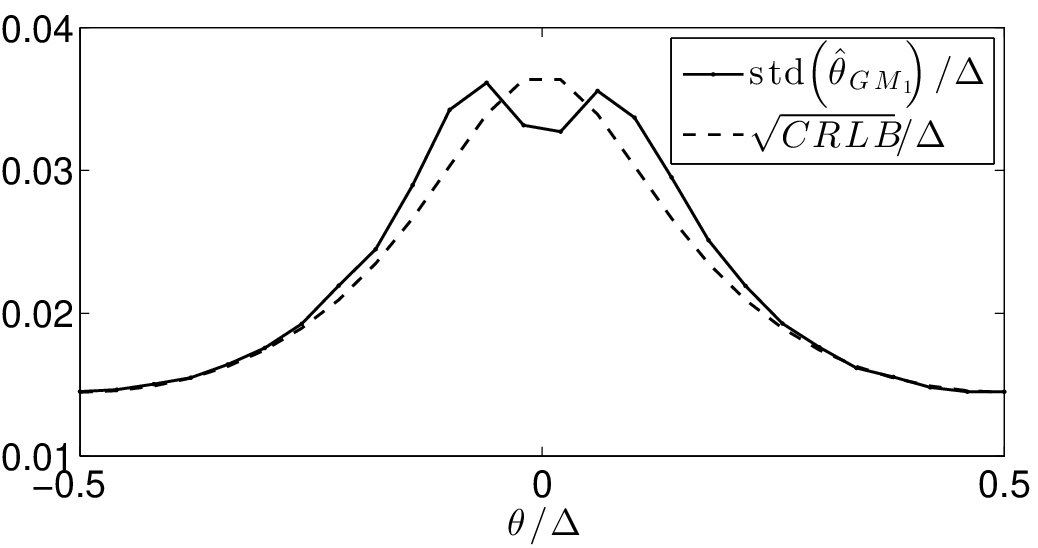}  
\caption{{\color{black} Model 1. Comparison between the
normalized standard deviation of $\hat{\theta}_{GM_1}$ obtained 
as in Fig.~\ref{errsone} and the corresponding Cramer--Rao lower bound for unbiased estimators, derived using an expression published in \cite{MoschittaSchoukensCarbone}. }
\label{errcramer}}
\end{center}
\end{figure}

Fig.~\ref{errmone}(c) shows that already with $\overline{\sigma}=0.2$ the estimation bias is largely removed in comparison to results shown in Fig.~\ref{errmone}(a). 
Fig.~\ref{errsone} shows the behavior of the normalized standard deviation of the estimation error.
Fig.~\ref{errsone}(a) refers to the arithmetic mean estimator while Fig.~\ref{errsone}(b)
and Fig.~\ref{errsone}(c) show the normalized standard deviation of $\hat{\theta}_{GM_1}$
when $\overline{\sigma}=0.17$ and $\overline{\sigma}=0.2$, respectively. 
Results show that in all cases the standard deviation remains of the same order of magnitude, with the newly proposed
estimator removing the largest part of the estimation bias.
{\color{black} Thus, the new estimator removes the bias at the expense of an increased variance. Since when $\overline{\sigma} > 0.4$, 
the arithmetic mean already shows negligible bias when the noise is Gaussian  \cite{Carbone}, the quantile--based estimator is more effective when the noise standard deviation is lower than this bound. Conversely, both estimators tend to provide similar results.}
Moreover, observe that already for $\overline{\sigma}=0.2$, the normalized standard deviation of 
$\hat{\theta}_{GM_1}$ approximately 
achieves the square--root of the 
Cramer--Rao lower bound applicable to unbiased estimators of a quantized constant in Gaussian noise
with known variance \cite{MoschittaSchoukensCarbone}. 
{\color{black} 
This becomes evident in Fig.~\ref{errcramer} where both the normalized
estimator variance and the corresponding Cramer--Rao lower bound are plotted 
as a function of $\theta/\Delta$, assuming $N=300$ and $\overline\sigma=0.2$.
While the two curves tend to coincide when $|\theta/\Delta| \simeq 0.5$, the small positive and negative differences for other values of $\theta/\Delta$ are due to the residual estimator bias. 
}
Thus, the proposed estimator is capable to be
statistically efficient under suitable values of its tuning parameters.

\begin{figure}[t!]
\begin{center}
\includegraphics[scale=0.35]{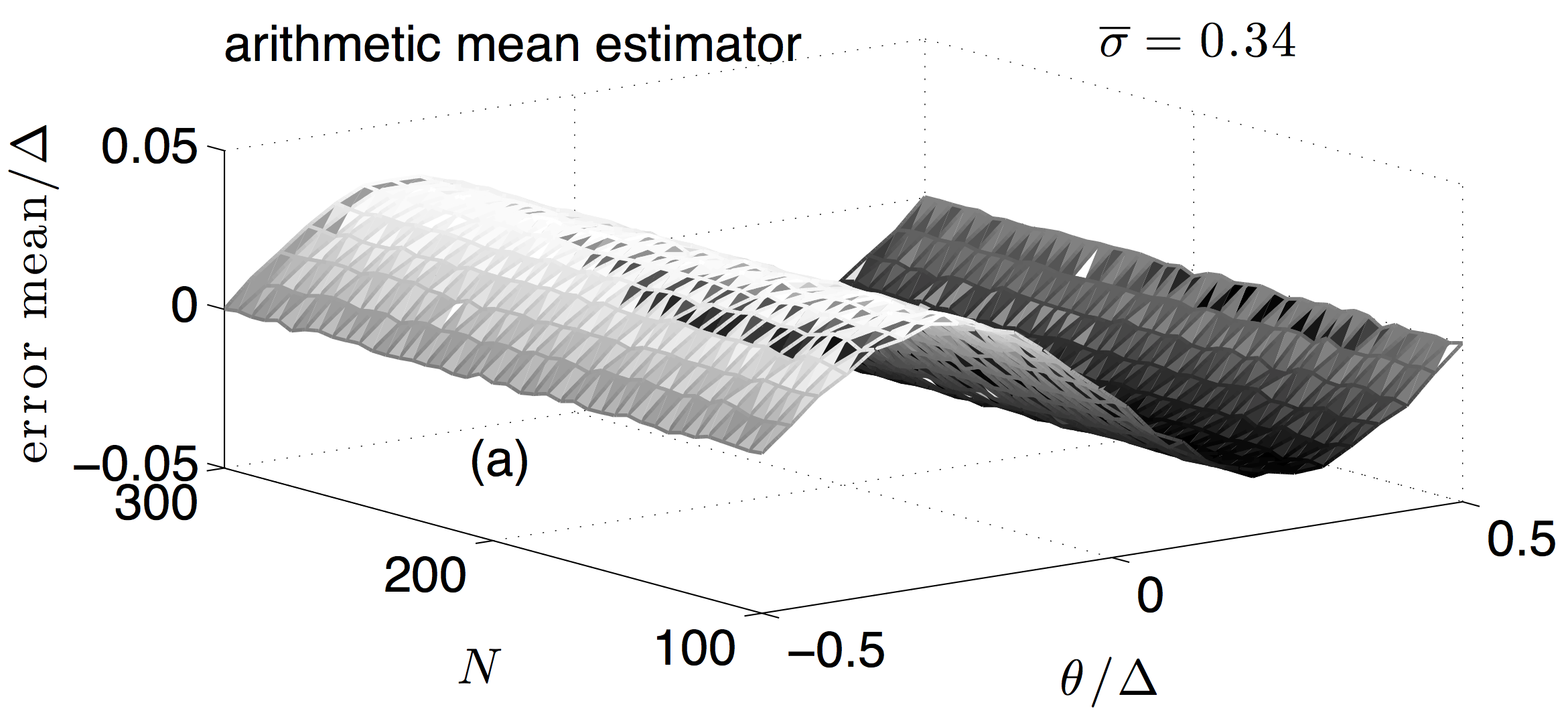}
\includegraphics[scale=0.35]{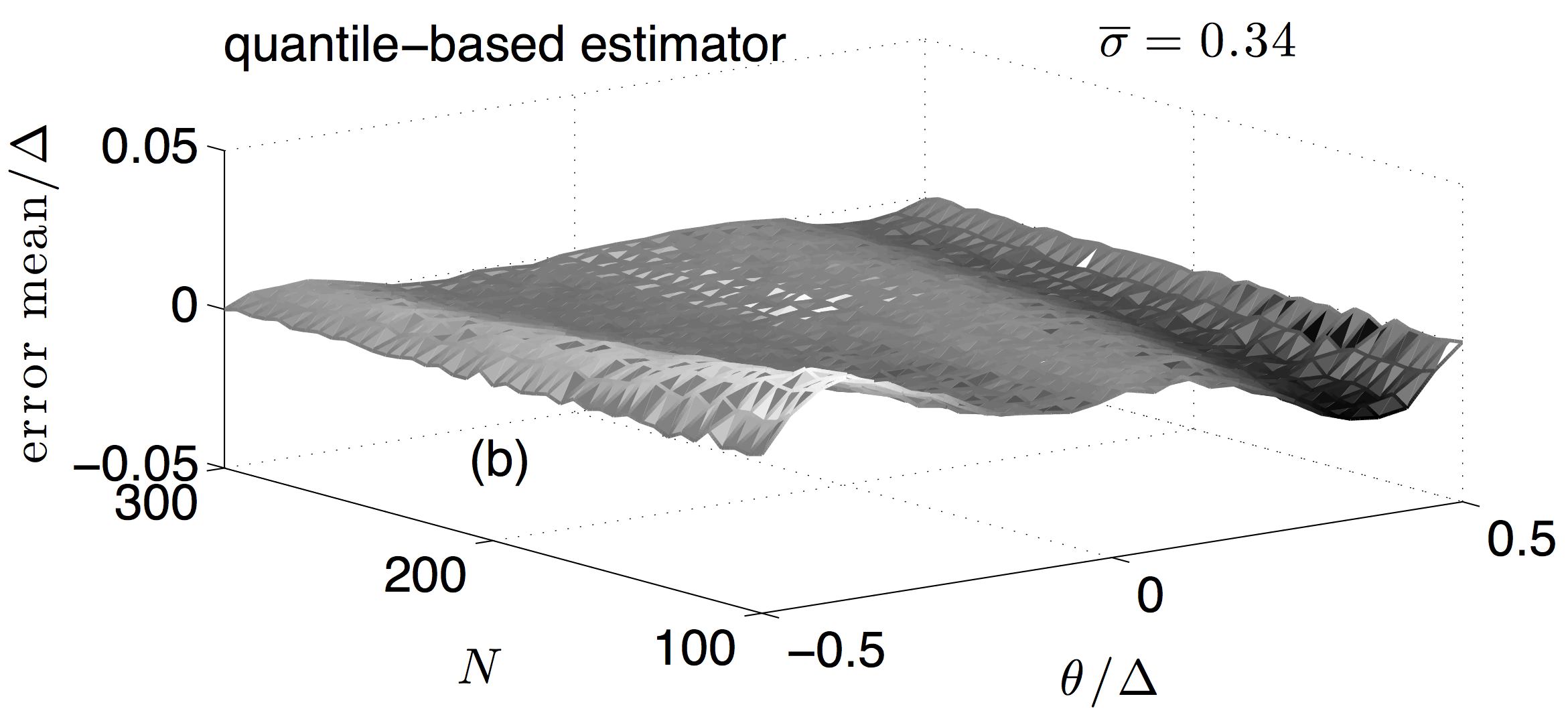}
\caption{Model 2. Mean value of the estimation error of a quantized constant in noise normalized to $\Delta$ (model~1). Montecarlo results based on $5000$ records as a function of $\nicefrac{\theta}{\Delta}$ and $N$, when $\overline{\sigma}=0.34$: arithmetic mean estimator (a), quantile--based estimator (b).  
\label{errtwo}}
\end{center}
\end{figure}

When model~2 is considered, the bias of the arithmetic mean estimator and that associated {\color{black} with} the quantile--based estimator are shown in Fig.~\ref{errtwo}(a) and (b), respectively. The noise standard deviation $\overline{\sigma}=0.34$ was assumed which results in an overall lower bias for the arithmetic mean estimator and its removal by the quantile--based estimator already when $N$ is slightly larger than $100$. Consider however that model~2 can be reduced to model~1 if the known value of the noise standard deviation is substituted by an estimate of it obtained using alternative estimators as in \cite{IEEE1241}\cite{IEEE1057}.

\begin{figure}[b!]
\begin{center}
\includegraphics[scale=0.47]{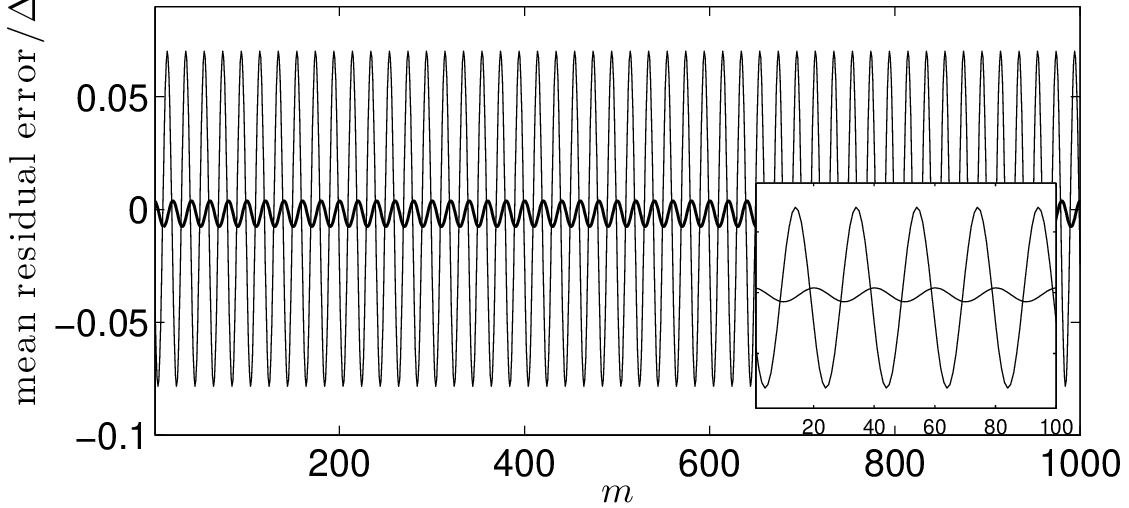}
\caption{Model 3. Mean value of the estimation error normalized to $\Delta$, as a function of the sampling index: the Gauss--Markov estimator (bold line) and the least--square estimator (thin line). Simulations were based on $1000$ Montecarlo records, known sequences $x_1[\cdot]$ and $x_2[\cdot]$ and $\sigma=0.3\Delta$, with $\Delta=2/2^{10}$. 
\label{errsinfit}}
\end{center}
\end{figure}

Simulations were also done when assuming model~3. Results are plotted in Fig.~\ref{errsinfit}, based on $1000$ Montecarlo records.
This figure shows the average residual error in the estimation of $s[\cdot]$ based on $1000$ records, when assuming  $x_1[n]=\cos\left(2\pi \frac{n}{M}N\right)$
and $x_2[n]=\sin \left(2\pi \frac{n}{M}N\right)$, with $M=20$ samples per period, $N=50$ periods and $\frac{\theta}{\Delta}=[3.7\; 11.4\; 23.1]$. A $10$
bit quantizer is considered having $\Delta=2/2^{10}$ and being affected by Gaussian noise with $\sigma=0.3\Delta$. 
For each time index $m$, the mean value of the difference between the simulated and estimated sinewaves is plotted in this figure.
To prove the validity of the new approach, the known quantization levels were assumed affected by INL uniformly distributed in $\left( -\Delta/2, \Delta/2 \right)$. 
Both the errors associated {\color{black} with} the usage of the proposed estimator (bold line) and of the well--known
least--square estimator (thin line) are shown.
The inset shows an enlarged view of the first $100$ samples.

Data in Fig.~\ref{errsinfit} show that the new estimator is capable of reducing the estimation bias and improving estimation accuracy. The new estimator requires knowledge of the transition level values, while the least--square estimator does not. While a calibration phase might be required in those cases where the ADC can not be considered to be linear enough, this additional information is exploited by the algorithm to improve estimation accuracy. Conversely, the least--square estimator does not include this information and its performance degrades progressively when the ADC behavior departs from ideality.

\begin{figure}[b!]
\begin{center}
\includegraphics[scale=0.35]{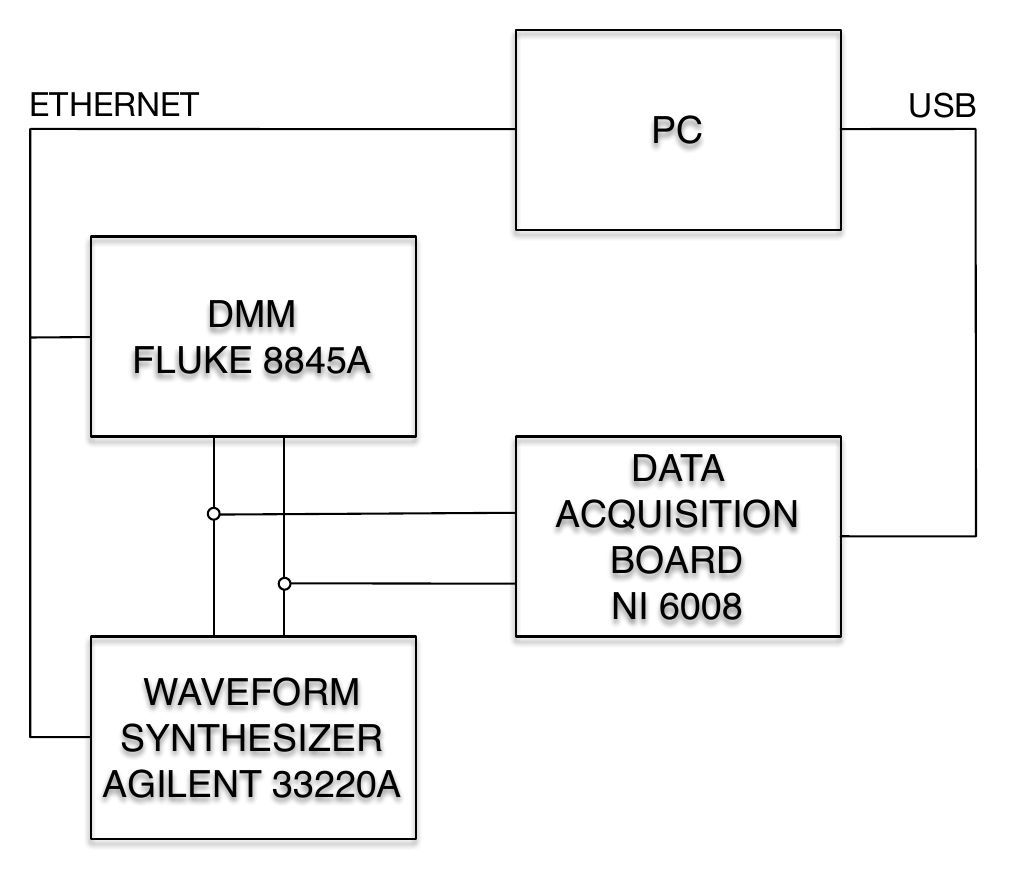}
\caption{Measurement setup used to collect experimental data.   
\label{expsetup}}
\end{center}
\end{figure}

\section{Experimental Results \label{expres}}
To prove the practical viability of the proposed estimator, the signal chain depicted in Fig.~\ref{expsetup}
was used. A $12$--bit commercial DAS with $\Delta=0.005085$~V
was used to prove the applicability of the proposed estimators.
The transition levels of the DAS, as defined in \cite{IEEE1241}, 
were first estimated, along with the DAS noise standard deviation $\hat \sigma_{DAS}\simeq 0.15\Delta$, 
measured as of clause 9.4.2 in \cite{IEEE1241}. 
The DAS was also calibrated for offset errors by simple pre--processing of all acquired data: raw data were multiplied by a gain and added to a constant value to remove
offset and the gain errors \cite{IEEE1241}.
Three sets of experiments were performed to test the proposed estimator under the hypotheses of {\color{black} model~1 and model~3}.

\subsection{Model~1 assumption: experimental results}
A waveform synthesizer used as a source of DC voltages affected by artificially added Gaussian noise, 
with noise standard deviation $\sigma_{\eta}$,
was used to provide input values to the DAS,
in the range $(-19 \Delta, 19 \Delta)$~V.
The total measured noise standard 
deviation $\sigma \simeq (\sigma_{\eta}^2+{\sigma}_{DAS}^2)^{\nicefrac{1}{2}}$, comprehensive of the 
DAS input--referred contribution, was estimated during the calibration phase as $\hat{\sigma} \simeq 0.264\Delta$. 

\begin{figure}[t!]
\begin{center}
\includegraphics[scale=0.44]{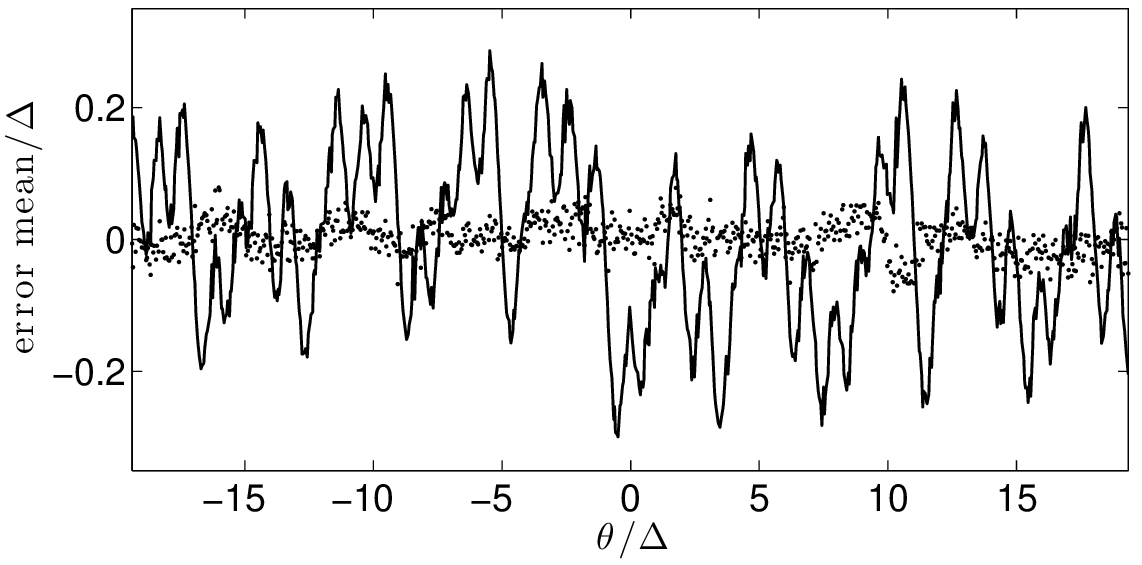}
\caption{Model 1. Mean value of the estimation error of a quantized constant in noise normalized to $\Delta$. 
Experimental results obtained using the setup shown in Fig.~\ref{expsetup}. and $N=500$.
Estimator bias as a function of $\nicefrac{\theta}{\Delta}$ when using 
the arithmetic mean estimator (solid line) and the quantile--based estimator (dots). 
For each $\nicefrac{\theta}{\Delta}$ a single record of $N=500$ was used.
Transition levels and input--referred noise standard deviation are first estimated using experimental data, 
for model~1 to be applicable (see text).   
\label{errresults}}
\end{center}
\end{figure}

A $6\nicefrac{1}{2}$--digit 
digital multimeter (DMM) was employed to measure a {\em true quantity value}  
of the DAS input \cite{VIM}. For each of the $800$ DC values in the input range a single record of $N=500$ acquisitions  
was collected by the DAS, along with the reference value measured by the DMM.
All instruments were connected to a PC using either the Ethernet or a USB connection.
Data were processed to obtain estimates of the applied DC value and of the corresponding 
estimation errors. Both the simple mean value estimator and the newly proposed estimator were used, 
under the assumptions of  model 1. Results are plotted in Fig.~\ref{errresults} using a solid line and dots, in the 
former and latter case, respectively.
They confirm the accuracy of the proposed procedure and show that
it is effective in removing the estimation bias characterizing the simple mean estimator.

\subsection{Model~3 assumption: experimental results}
\begin{figure}[t!]
\begin{center}
\includegraphics[scale=0.44]{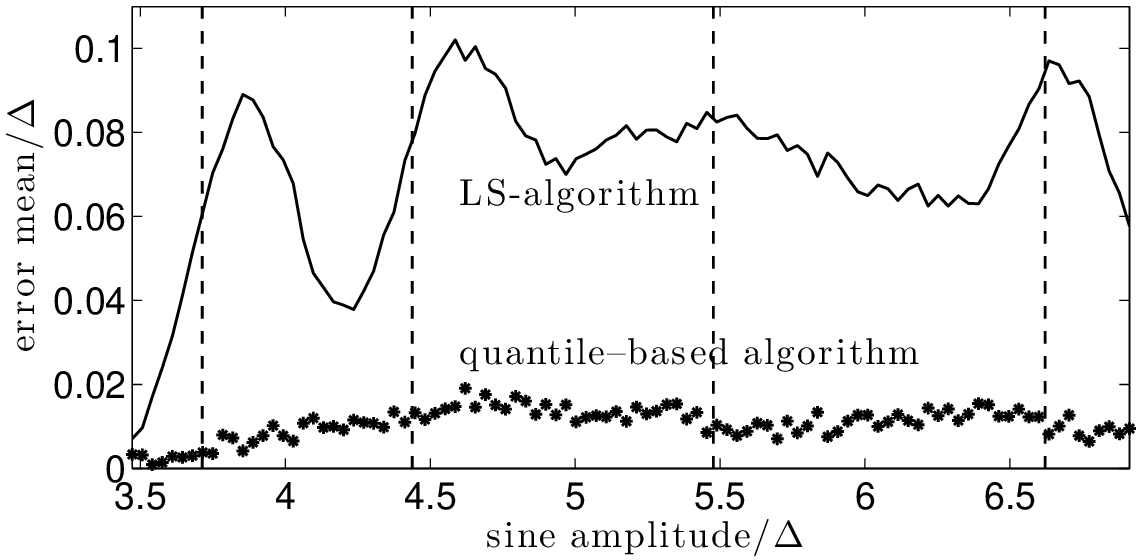}
\caption{Model 3. Mean value of the estimation error of the amplitude of a synchronously sampled sinewave
normalized to $\Delta$, based on $5$ records of $8000$ points each.
Experimental results obtained using the setup shown in Fig.~\ref{expsetup}.
Estimator bias as a function of $100$ sinewave amplitude in the interval $(0.05,0.1) \;$V,  when using 
the least--square estimator (solid line) and the quantile--based estimator (stars). 
ADC transition levels, whose value is shown using dashed lines, and input--referred noise standard deviation are first estimated using experimental data, for model~3 to be applicable (see text).   
\label{ACerrresults}}
\end{center}
\end{figure}

A waveform synthesizer was used to generate a $100\;$ Hz sinewave with $100$ variable amplitudes in the range $(0.05,0.1)\;$ V,
without the addition of noise. This signal was acquired by the DAS sampling at $10^4$ ksample/s, providing $M=100$ samples per 
period.
For each one of the $100$ voltage values, $8000$ points resulting in $N=80$ periods, were recorded and processed using both the standard
least--square method described in \cite{IEEE1241} and the newly proposed quantile--based estimator to estimate $\theta_1, \theta_2, \theta_3$. The sinewave amplitude was then estimated as $\sqrt{\hat{\theta}_1^2+\hat{\theta}_2^2}$ , with 
$\hat{\theta}_i$, $i=1,2$ as each one of the two estimators. 
The amplitude reference value was measured by the DMM put in AC mode, as the mean value of $10$ measurement results. 
The mean error obtained using $5$ records for each sinewave amplitude is plotted in Fig.~\ref{ACerrresults} in both cases,
using a solid line for the LS--estimator and stars for the newly proposed estimator. 
Given that only the sinewave amplitude is estimated, there was no need to synchronize the DAS with the waveform synthesizer.
Thus, the initial record phase of the sinewave generator
was not controlled and was assumed as a uniform random variable in $[0, 2\pi)$.

\balance

\section{Discussion of Results}
{\color{black}
In this Section, at first general properties of the proposed estimator are discussed on the basis of the described
results. Then, known estimator issues and 
corresponding fixes are examined.
\subsection{Performance comparison}
}
The proposed estimator can be used to obtain accurate measurements of parameters of
DC and AC sequences if some of the characteristic parameters
of the DAS used to quantize data are known, such as transition levels and input--referred noise standard deviation. 
Eventual errors in the estimation of these parameters performed beforehand, will not affect significantly the 
performance of the estimator, as shown by data in Fig.~\ref{errresults} and \ref{ACerrresults}.
If nominal thresholds values are used instead of actual values, the estimator still provides reasonable results,
improving over the performance of the simple mean estimator. 
Provided that the noise standard deviation is large enough to excite a number of quantization bins exceeding at least by one the number of parameters to be estimated, there are no restrictions on the severity of the applied quantization:
accurate estimates are obtained even when data are quantized using low--resolution DASs.

In practice, the effect of noise is that of encoding the information about the unknown parameters' 
values in the probabilities with which the various
quantizer output codes occur. The algorithm presented here acts as a decoder of such information, by properly processing the estimated probabilities. Moreover, while the usual approach to process quantized data is based, at most, on calibration
of ADCs for offset and gain, the presented estimator allows calibration at the transition level and is thus {\em inherently}
more statistically powerful. 
Finally, the same approach followed here can be applied with noise PDFs different from the Gaussian one,
by recalculating the covariance matrix and by using the proper quantile equation in (\ref{linearmodel}) and in (\ref{linearmodell}).  
{\color{black}
\subsection{Known issues and fixes}
It is not assured that the quantile--based estimator uniformly outperforms other estimators. For instance, consider the case shown in Fig.~\ref{figfail}, where the mean--square error is plotted,
under model~1, for both the quantile--based and the arithmetic mean estimators. Here, $N=1000$ is assumed and $1000$ records
of $\theta$ varying in the interval $[-0.1 \Delta, 0.1\Delta]$
are considered. By knowing that the closest transition levels are positioned in $-\nicefrac{\Delta}{2}$ and $\nicefrac{\Delta}{2}$, 
it can be observed that when 
$\theta \simeq 0$ the quantile--based estimator maintains its overall behavior. Conversely, 
the arithmetic mean estimator benefits from the 
DC level getting close to an ADC equivalent output level ($\simeq 0$) and thus outperforms the quantile--based estimator.

Moreover, the proposed quantile--based estimator may fail 
when the record length $N$ is so small 
or  $\overline{\sigma} \ll 1$, so 
that all samples in the record belong to the same quantization bin.
This results in a situation where all collected samples excite the same bin that would be excited if the noise would not be present.
Under model~1 this results in
$\Lambda=0$ 
and the estimator not having enough information to identify
the model.  In a similar way model~2 and model~3 may not be identifiable if $\Lambda$ drops below $2$ and $3$, respectively.
These events are unlikely if $N \gg 1$ and/or $\overline{\sigma} > \nicefrac{1}{6}$. 
However, an easy fix could be proposed by resorting to a mechanism that recognizes the occurrence of such phenomena and switches to more conventional estimators, such as the LSE or the MLE.
\begin{figure}[t!]
\begin{center}
\includegraphics[scale=0.5]{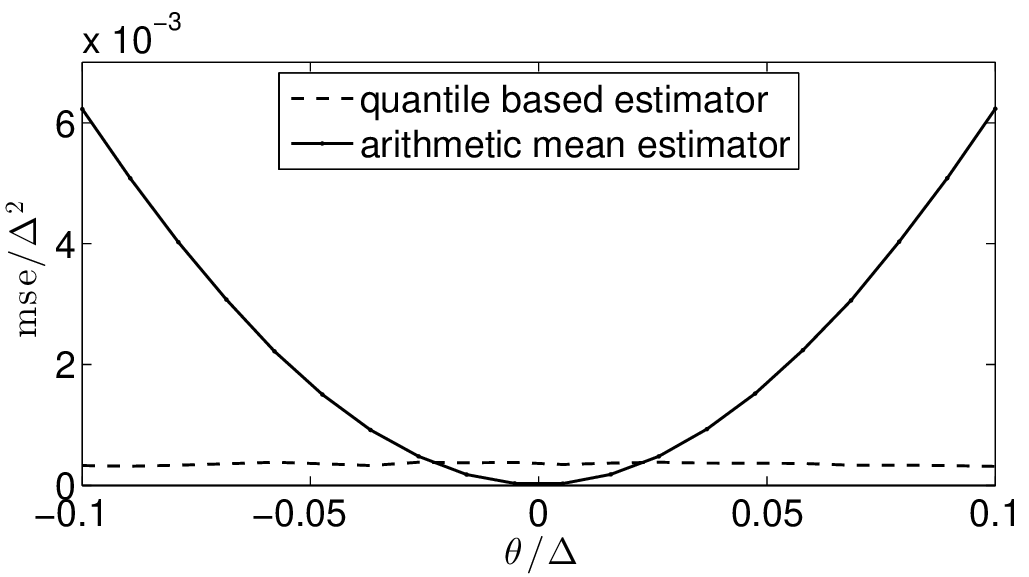}
{\color{black}
\caption{Model 1. Mean--square error in the
 the estimation of a quantized constant in noise normalized to $\Delta^2$. Montecarlo results based on $1000$ records of $N=1000$ samples each, as a function of $\nicefrac{\theta}{\Delta}$, when $\overline{\sigma}=0.2$.
\label{figfail}}
}
\end{center}
\end{figure}
}

\section{Conclusion}
An estimator is presented in this paper that exploits the unbiasedness of a quantile estimator based on ADC output data, when 
the quantile level coincides with {\color{black}the value of one of} the ADC transition levels. The estimator is based on a linear relationship between 
a statistics of the observed data using nonlinear functions and parameters to be identified, 
and is thus easily computable using matrix calculations. 
By also taking into account the covariance between the 
ADC output codes, 
it was possible to show that this estimator is an application of the Gauss--Markov theorem and that it is rather robust toward inaccuracies in some of the necessary hypotheses.

\section*{Acknowledgement}
This work was supported in part by the Fund for Scientific Research (FWO-Vlaanderen), by the Flemish Government (Methusalem), the Belgian Government through the Inter university Poles of Attraction (IAP VII) Program, and by the ERC advanced grant SNLSID, under contract 320378.


\begin{thebibliography}{12}
\bibitem{Carbone}P.~Carbone, ``Quantitative criteria for the design of dither-based quantizing systems,'' {\em IEEE Trans. Instr. 
Meas.}, vol. 46, no. 3, pp. 656--659, June 1997.
\bibitem{Kollarbias}I.~Koll\'ar, ``Bias of mean value and mean square value measurement based on quantized data,'' {\em IEEE Trans. Instr. Meas.}, vol. 43, pp. 373--379, Oct. 1994.
\bibitem{Schuchman}L.~Schuchman,  ``Dither signals and their effect on quantization noise,'' {IEEE Transaction on Communication Technology}, vol. 12, pp. 162--165, 1964.
\bibitem{GrayStockham} R.~M.~Gray, T.~G.~Stockham, ``Dithered Quantizers,'' {\em IEEE Trans. Inform. Theory}, Vol. 39, no. 3, May 1993, pp. 805--812.
\bibitem{GrayNeuhoff} R.~M.~Gray, D.~L.~Neuhoff, ``Quantization,'' {\em IEEE Trans. Inform. Theory}, Vol. 44, no. 6, Oct. 1998, pp. 2325--2383.
\bibitem{KollarBook}B.~Widrow and I.~Koll\'ar, {\em Quantization Noise},  Cambridge University Press, 2008.
\bibitem{Giaquinto} A.~Di~Nisio, L.~Fabbiano, N.~Giaquinto, M.~Savino ``Statistical Properties of an ML Estimator for Static ADC Testing,'' Proc. of 12th IMEKO Workshop on ADC Modelling and Testing, Iasi, Romania, Sept. 19--21, 2007.
\bibitem{Gendai} Y.~Gendai, ``The Maximum--Likelihood Noise 
Magnitude Estimation in ADC Linearity Measurements," {\em IEEE Trans. Instr. Meas.}, July 2010, vol. 59, no. 1, pp. 1746--1754.
\bibitem{MoschittaSchoukensCarbone}A. Moschitta, J.~Schoukens, P.~Carbone, ``Information and Statistical Efficiency
When Quantizing Noisy DC Values,'' accepted for publication in {\em IEEE Trans. Instr. Meas.}.
\bibitem{Kollar1}
L. Balogh, I. Koll\'ar, L. Michaeli, J. \v{S}aliga, J. Lipt\'ak,
``Full information from measured ADC test data using maximum likelihood estimation,'' {\em Measurement}, vol. 45, pp. 164--169, 2012.
\bibitem{Kollar2}
J. \v{S}aliga, I. Koll\'ar, L. Michaeli, J. Bu\v{s}a, J. Lipt\'ak, T. Virosztek, 
``A comparison of least squares and maximum likelihood methods using sine fitting in ADC testing,'' {\em Measurement}, vol. 
46, pp. 4362--4368, 2013.
\bibitem{Gustafsson1} F.~Gustafsson, R.~Karlsson, ``Generating dithering noise for maximum likelihood estimation from quantized data,'' {\em Automatica}, Vol. 49, pp. 554--560, 2013.
\bibitem{WangYinZhangZhao}L. Y. Wang, G. G. Yin, J. Zhang, Y. Zhao, {\em  System Identification with Quantized Observations,} Springer Science, 2010.
\bibitem{104} L.~Y.~Wang and G.~Yin, ``Asymptotically efficient parameter estimation using quantized output observations,'' {\em Automatica}, Vol. 43, 2007, pp. 1178--1191.
\bibitem{IEEE1241} IEEE, {\em Standard for Terminology and Test Methods for Analog--to--Digital Converters}, IEEE Std. 1241, Aug. 2009.
\bibitem{multinomial}M.~Evans, N.~Hastings, B.~Peacock, {\em Statistical Distributions -- 3rd ed.}. Wiley, New York, USA 2000. 
\bibitem{Knapp}T.~R.~Knapp, ``Treating Ordinal Scales as Interval Scales: An Attempt To Resolve the Controversy,'' {\em Nursing Research}, March/April, vol. 39, no. 2, 1990.
\bibitem{Stevens}S.~S.~Stevens, ``On the Theory of Scales of Measurement,'' {\em Science}, vol. 103, no. 2684, June 1946.
\bibitem{Kay}S.~M.~Kay, {\em Fundamentals of Statistical Signal Processing,} Prentice--Hall, 1998.
{\color{black}
\bibitem{Kai}K. O. Arra, ``An Introduction To Error Propagation: Derivation, Meaning and Examples of Equation $C_Y = F_X C_X F_X^T$," {\em Technical Report}, Swiss Federal Institute of Technology Lausanne, 1998. [online] http://www.nada.kth.se/\textasciitilde kai-a/papers/arrasTR-9801-R3.pdf.} 
\bibitem{IEEE1057}IEEE, {\em Standard for Terminology and Test Methods for Waveform Digitizers}, IEEE Std. 1057, Aug. 2007.
\bibitem{VIM} JCGM 200:2012, {\em International vocabulary of metrology -- Basic and general concepts and associated terms, 3rd ed.}, (VIM), ISO, BIPM, IEC, IFCC, ISO, IUPAC, IUPAP and OIML, 2012.  [online] http://www.bipm.org/utils/common/documents/jcgm/JCGM\_200\_2012.pdf.
\bibitem{CarboneVandersteen}P.~Carbone, G.~Vandersteen, ``Bias Compensation When Identifying Static Nonlinear Functions Using  Averaged Measurements," {\em IEEE Trans. Instr. Meas.}, 2014.

\end{thebibliography}
\end{document}